\documentclass[twoside]{article}
\usepackage[sc]{mathpazo}
\usepackage[T1]{fontenc}
\usepackage{microtype}
\usepackage[hmarginratio=1:1,top=32mm,columnsep=20pt]{geometry}
\usepackage{multicol}
\usepackage[hang, small,labelfont=bf,up,textfont=it,up]{caption}
\usepackage{booktabs}
\usepackage{float}
\usepackage{hyperref}
\usepackage{amsmath}
\usepackage{authblk}
\usepackage{lettrine}
\usepackage{paralist}
\usepackage{pgfplots}
\usepackage{abstract}
\usepackage{titlesec}
\usepackage{longtable}
\usepackage{natbib}
\usepackage{algorithm,algorithmic}
\usepackage{subcaption}

\bibpunct{(}{)}{;}{a}{}{;}

\usetikzlibrary{positioning}
\usetikzlibrary{fit}
\usetikzlibrary{backgrounds}
\usetikzlibrary{calc}
\usetikzlibrary{shapes}
\usetikzlibrary{mindmap}
\usetikzlibrary{decorations.text}
\usetikzlibrary{shapes,arrows}

\renewcommand\thesection{\arabic{section}}
\renewcommand\thesubsection{\Alph{subsection}}

\titleformat{\section}[block]{\large\scshape\centering}{\thesection.}{1em}{}
\titleformat{\subsection}[block]{\large}{\thesubsection.}{1em}{}

\title{\vspace{-15mm}\fontsize{13pt}{10pt}\selectfont\textbf{The Problem of Calibrating an Agent-Based Model of High-Frequency Trading}}

\author[1,2]{D. F. Platt \thanks{Corresponding author, donovan.platt@students.wits.ac.za}}
\author[1,2]{T. J. Gebbie}
\affil[1]{\small School of Computer Science and Applied Mathematics, University of the Witwatersrand, Johannesburg, South Africa}
\affil[2]{\small QuERILab - Quantifying Emergence, Risk and Information}

\date{}

\begin{document}

\maketitle


\vspace{-1cm}

\begin{abstract}

\noindent Agent-based models, particularly those applied to financial markets, demonstrate the ability to produce realistic, simulated system dynamics, comparable to those observed in empirical investigations. Despite this, they remain fairly difficult to calibrate due to their tendency to be computationally expensive, even with recent advances in technology. For this reason, financial agent-based models are frequently validated by demonstrating an ability to reproduce well-known log return time series and central limit order book stylized facts, as opposed to being rigorously calibrated to transaction data. We thus apply an established financial agent-based model calibration framework to a simple model of high- and low-frequency trader interaction and demonstrate possible inadequacies of a stylized fact-centric approach to model validation. We further argue for the centrality of calibration to the validation of financial agent-based models and possible pitfalls of current approaches to financial agent-based modeling.

\vspace{0.3cm}


\vspace{0.3cm}


\end{abstract}


\begin{multicols}{2}

\section{Introduction}

Agent-based modeling is a relatively new technique that has shown increased prevalence in recent years, with applications across a diverse array of fields \citep{Macal10}. Agent-based models (ABMs) are characterized by a bottom-up development process, in which the properties of and interactions between the micro-constituents of a system are modeled as opposed to the overall system dynamics \citep{Macal10}. 

The appeal of these models can be strongly linked to the increased focus on the discipline of complexity science in contemporary literature, in which it is argued that many systems exhibit emergent properties which cannot be explained by the properties of their constituent parts, but rather result from the interactions between these constituent parts \citep{Heylighen08}. \citet{Macal10} indicate that this focus on representing individual agents and their interactions within ABMs leads to emergent behaviors not explicitly programmed, demonstrating an ability to model phenomena which could be described as complex. 

One should, however, be skeptical of the idea that all complex behavior in a system can be replicated using a bottom-up modeling approach. \citet{Wilcox14} propose the idea of hierarchical causality in the financial system, stating that there is in fact a hierarchy representing different levels of abstraction, each leading to different nuances in overall behavior. This would imply that there is both bottom-up and top-down causation within the financial system and that replicating complex behavior within it is more nuanced than simply replicating the bottom-up processes that may be present. Nevertheless, financial ABMs represent a significant attempt at the replication of complex behaviors emerging in financial systems.

Considering the fact that overall system behavior is not explicitly modeled in most cases, ABMs dispense with most assumptions regarding overall system dynamics when compared to other approaches, particularly relevant in financial applications where common assumptions, such as the Gaussianity of log return distributions, market efficiency and rational expectations have become increasingly criticized due to incompatibility with empirical observations \citep{LeBaron05}. Not surprisingly, the absence of these assumptions has led to the existence of a wide variety of financial ABMs, all able to replicate empirically-observed return time series stylized facts, such as volatility clustering and a fat-tailed distribution \citep{Barde16}.

Despite the apparent successes of financial agent-based modeling, there still exist many critics of the approach, particularly skeptical of the extent to which the models have been validated \citep{Hamill16}. \citet{Fabretti13} notes that the literature regarding the application of quantitative calibration techniques to financial ABMs remains relatively sparse, with much of the remaining literature relying on less rigorous qualitative calibration techniques, in which parameters are manually selected such that well-known log return time series stylized facts are recovered by the candidate models. A similar approach is also employed in models including central limit order book (LOB) stylized facts, such as in the work conducted by \citet{Preis06}.

While the recovery of stylized facts is a good starting point to indicate that a model may be based upon sound principles, rigorous validation would require a model to produce return time series with statistical properties similar to those observed in actual transaction data, such that a simulated series could be said to come from the same distribution as a series drawn from empirical data \citep{Fabretti13}. We thus require model parameters that allow a particular ABM to reproduce these statistical properties. This, in turn, results in a calibration problem in which the model must be calibrated to transaction data to obtain these parameters. There is thus an intrinsic link between calibration and validation. 

It should be noted that the vast majority of financial ABM calibration literature involves the application of calibration techniques to models involving closed-form solutions for prices applicable at a daily time scale. In essence, prices are determined as weighted sums or averages of individual trader price expectations or order prices and sizes during each iteration of a simulation representing a series of trading sessions, with a session typically representing a single day of trading. Members of this class of models, representing earlier work in the field, include the \citet{Farmer02} and \citet{Kirman91} models, among many others.

This simplification is generally appropriate for the daily time scale considered by these models, since this mechanism could be seen as analogous to the closing auctions found at the end of each trading day in real financial markets, where trader demands, represented as unexecuted limit orders, are considered and a closing price determined which maximizes the executable volume in the closing auction period. This would typically involve a form of averaging. It is clear, however, that such a simplification would not be appropriate for intraday time scales and hence high-frequency trader activity.

A number of successful strategies have emerged for the calibration of models with closed-form solutions for prices, with the two most prominent being the method of simulated moments and maximum likelihood estimation. The work of \citet{Gilli03} and \citet{Fabretti13}, on which we base our investigation, details the application of the method of simulated moments. Similar experiments involving maximum likelihood estimation are presented in \citet{Alfarano05}, \citet{Alfarano06} and \citet{Alfarano07}. Also worth noting is the work of \citet{Amilon08}, which presents a comparison of maximum likelihood and the efficient method of moments.

Despite the pervasiveness of the above two methods in ABM calibration literature, new methods and augmentations to existing approaches have also been proposed in recent years, with computational constraints remaining a key obstacle. Notable examples include the work of \citet{Recchioni15}, which employs gradient methods, the work of \citet{Guerini16}, which estimates vector autoregressive models from simulated and real world data and compares their structure, and the work of \citet{Grazzini15a}, which presents a discussion on approaches able to estimate ergodic models \citep{Grazzini12} and in what circumstances such approaches are successful. A fairly exhaustive survey of existing ABM calibration methods is presented by \citet{Kukacka16}. 

In contrast to models with closed-form solutions for prices, contemporary models, such as those of \citet{Preis06}, \citet{Chiarella09} and \citet{JacobLeal15}, focus instead on recreating double auction markets at an intraday time scale in which agent orders are stored in a central LOB and orders are executed using realistic matching processes. It is apparent, however, that the addition of realistic matching processes severely increases the computational expense associated with the simulation process, which may have an adverse effect on calibration attempts. Furthermore, many of the previously mentioned calibration methodologies tend to assume and therefore require a closed-form solution for prices. These are likely reasons why models of this class still remain largely uncalibrated in a convincing manner.

It is important to be aware of what the idealistic sufficient and necessary conditions are for a model to be considered identifiable, namely the objective function having a reasonable and unique extremum at the true parameter values \citep{Canova09}. Some caution is required in the context of ABMs of the type we consider as they can at best be approached via indirect estimation to provide consistency between moments and data as opposed to being rigorously estimated; hence we retain the terminology of calibration \citep{Cooley97} as opposed to estimation, in the sense that we attempt to provide consistency between the data and model parameters because we cannot directly test the structure of the models themselves.

The primary focus of this investigation is to expand upon the work conducted by \citet{Fabretti13}, who introduces a strategy for the calibration of the \cite{Farmer02} model, one of the aforementioned daily models with a closed-form solution for prices. Using this strategy, we attempt to calibrate the \citet{JacobLeal15} model, a more sophisticated intraday ABM representing a double auction market based around a central LOB, with both high- and low-frequency trader activity present. A key advantage of the method of \citet{Fabretti13} is that it does not require the candidate model to have a closed-form solution for prices, making it compatible with the \citet{JacobLeal15} model.

As per the investigation conducted by \citet{Fabretti13}, our approach involves the construction of an objective function using the method of simulated moments and a moving block bootstrap \citep{Winker07}, followed by the application of heuristic optimization methods, namely the Nelder-Mead simplex algorithm with threshold accepting \citep{Gilli03} and genetic algorithms \citep{Holland75}. The need for heuristic optimization methods stems from the tendency of objective functions in our context to lack smoothness, which causes traditional optimization methods to find local as opposed to global minima \citep{Gilli03}.

The dataset used in calibration consists of Thomson Reuters Tick History (TRTH) data for a stock listed on the Johannesburg Stock Exchange (JSE). Prior to a change in the billing model towards the end of 2013, the JSE had minimum transaction costs that made high-frequency trading largely unprofitable \citep{Harvey17}. We thus apply the aforementioned calibration methodology in an attempt to gain a deeper understanding of the emergence of high-frequency trading in a developing market.

\section{The Jacob Leal et al. Model \label{Model}}

Referring to Figure \ref{ModelOverview}, the model consists of two agent types, high-frequency (HF) and low-frequency (LF) traders, where each LF trader may be following either a chartist or fundamentalist strategy. These agents interact through the posting of limit orders to a central LOB, with time progressing through a series of $T$ one-minute trading sessions. 

During each session, each active trader agent places a single limit order, with price and size determined by the trader's current strategy. Thereafter, limit orders in the order book are matched according to price and then time priority until all possible trades have been executed, with the final trade price defining the market price for the session. 

\tikzstyle{HF} = [rectangle, draw, fill=blue!20, text width=5em, text centered, rounded corners, minimum height=4em]
\tikzstyle{LF} = [rectangle, draw, fill=red!20, text width=5em, text centered, rounded corners, minimum height=4em]
\tikzstyle{process} = [rectangle, draw, fill=green!20, text width=5em, text centered, rounded corners, minimum height=4em]
\tikzstyle{terminal} = [circle, draw, fill=green!20, text width=5em, text centered, rounded corners, minimum height=4em]
\tikzstyle{line} = [draw, -latex']

\begin{figure}[H]

\begin{center}

\scalebox{0.6}{
\begin{tikzpicture}[node distance = 4.5cm, auto] 
\node [process] (LOB) {LOB};
\node [LF, right of=LOB] (tradeLF) {LF Traders (Chartist, Fundamentalist)};
\node [HF, left of=LOB] (tradeHF) {HF Traders};
\node [process, below of = LOB] (price) {Market Price Time Series};
\path [line, align=left] (tradeLF) -- node [midway, below, scale=0.8] {Limit Orders} (LOB);
\path [line, align=left] (tradeHF) -- node [midway, below, scale=0.8] {Limit Orders} (LOB);
\path [line] (LOB) edge[loop above] node [midway, above, scale=0.8] {Match Orders, Calculate Prices} ();
\path [line, align=left] (LOB) -- node [midway, right, scale=0.8] {Results in} (price);
\end{tikzpicture}
}

\end{center}

\caption{An illustration of the agent types and interactions within the model \label{ModelOverview}}

\end{figure}
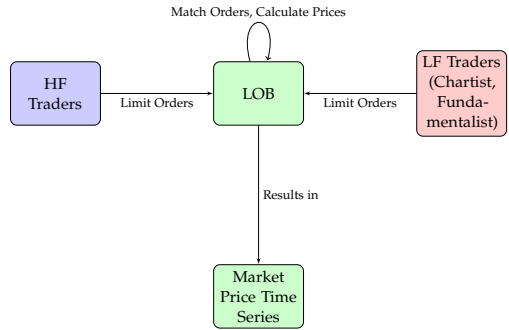

\tikzstyle{HF} = [rectangle, draw, fill=blue!20, text width=5em, text centered, rounded corners, minimum height=4em]
\tikzstyle{LF} = [rectangle, draw, fill=red!20, text width=5em, text centered, rounded corners, minimum height=4em]
\tikzstyle{process} = [rectangle, draw, fill=green!20, text width=5em, text centered, rounded corners, minimum height=4em]
\tikzstyle{terminal} = [circle, draw, fill=green!20, text width=5em, text centered, rounded corners, minimum height=4em]
\tikzstyle{line} = [draw, -latex']

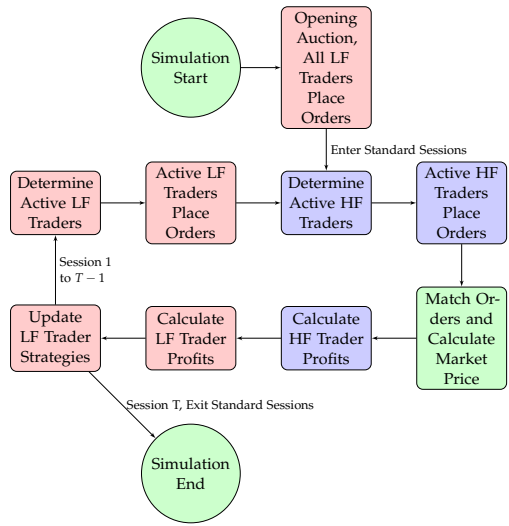
\begin{figure}[H]

\begin{center}

\scalebox{0.6}{
\begin{tikzpicture}[node distance = 3cm, auto] 
\node [LF] (activeLF) {Determine Active LF Traders};
\node [LF, right of=activeLF] (tradeLF) {Active LF Traders Place Orders};
\node [HF, right of=tradeLF] (activeHF) {Determine Active HF Traders};
\node [HF, right of=activeHF] (tradeHF) {Active HF Traders Place Orders};
\node [process, below of=tradeHF] (match) {Match Orders and Calculate Market Price};
\node [HF, left of=match] (profitHF) {Calculate HF Trader Profits};
\node [LF, left of=profitHF] (profitLF) {Calculate LF Trader Profits};
\node [LF, left of=profitLF] (strategyLF) {Update LF Trader Strategies};
\node [LF, above of=activeHF] (openAuction) {Opening Auction, All LF Traders Place Orders};
\node [terminal, left of=openAuction] (start) {Simulation Start};\
\node [terminal, below of=profitLF] (end) {Simulation End};
\path [line] (activeLF) -- (tradeLF);
\path [line] (tradeLF) -- (activeHF);
\path [line] (activeHF) -- (tradeHF);
\path [line] (tradeHF) -- (match);
\path [line] (match) -- (profitHF);
\path [line] (profitHF) -- (profitLF);
\path [line] (profitLF) -- (strategyLF);
\path [line] (start) -- (openAuction);
\path [line, align=left] (openAuction) -- node [midway, right, scale=0.8] {Enter Standard Sessions} (activeHF);
\path [line, align=left] (strategyLF) -- node [midway, right, scale=0.8] {Session 1 \\ to $T - 1$} (activeLF);
\path [line, align=left] (strategyLF) -- node [midway, right, scale=0.8] {Session T, Exit Standard Sessions} (end);
\end{tikzpicture}
}

\end{center}

\caption{Flowchart describing a typical simulation of T one-minute trading sessions \label{ModelEvents}}

\end{figure}

It should be noted that at any point in the simulation, all traders have access to the history of past market prices and fundamental values as well as the current state of the LOB.

In Figure \ref{ModelEvents}, we summarize the key events associated with \citet{JacobLeal15} model trading sessions. The opening auction is our own added initialization process, discussed in detail in Section \ref{Model}.\ref{Modifications}.

\subsection{LF Trader Agent Specification}

Each simulation contains $N_L$ LF traders, where different LF traders are represented by the index $i = 1,...,N_L$. Each LF trader has its own trading frequency, $\theta _i$, drawn from a truncated exponential distribution with mean $\theta$ and upper and lower bounds $\theta_{max}$ and $\theta_{min}$ respectively. This is described by \citet{JacobLeal15} as representing the chronological time frame associated with LF traders. This time frame is discussed at length by \citet{Easley12}. At each activation, active LF traders randomly select either the chartist or fundamentalist strategy for the session, with the probability of being chartist given by $\Phi_{i, t}^{c}$.

We now elaborate on the simulation event overview presented in Figure \ref{ModelEvents} with appropriate equations for an arbitrary session $t \leq T$. 

After activation, the i-th LF trader places a limit order, which requires the specification of both an order size and an order price. Should the i-th LF trader agent be following the chartist strategy, we calculate the order size according to
\begin{equation}
D_{i,t}^{c} = \alpha^{c}(\bar{P}_{t-1} - \bar{P}_{t - 2}) + \epsilon_{t}^{c} \label{Chartist}
\end{equation}
where $\bar{P}_{t}$ is the market price (in units of currency) for session $t$, $0 < \alpha^c < 1$ and $\epsilon_{t}^{c} \sim \mathcal{N}\left(0, (\sigma^c)^2\right)$. 

Conversely, in the case of the fundamentalist strategy, we calculate the order size according to
\begin{equation}
D_{i,t}^{f} = \alpha^{f}(F_{t} - \bar{P}_{t - 1}) + \epsilon_{t}^{f} \label{Fundamentalist}
\end{equation}
where $F_{t} = F_{t-1}(1 + \delta)(1 + y_{t})$ is the fundamental price (in units of currency) for session $t$, $\delta > 0$, $0 < \alpha^f < 1$, $\epsilon_{t}^{f} \sim \mathcal{N}\left(0, (\sigma^f)^2\right)$ and $y_{t} \sim \mathcal{N}\left(0, (\sigma^y)^2\right)$. 

The order price is identical under both strategies and is determined according to
\begin{equation}
P_{i, t} = \bar{P}_{t - 1}(1 + \delta)(1 + z_{i, t}) \label{LFOrderPrice}
\end{equation}
where $z_{i,t} \sim \mathcal{N}\left(0, (\sigma^z)^2\right)$. 

Once the order is placed, it remains in the LOB until it is matched or until $\gamma^L$ sessions have passed, resulting in its removal from the order book. At the end of each session, the i-th LF trader calculates their profits according to
\begin{equation}
\pi_{i, t} ^ {s} = (\bar{P}_{t} - P_{i, t})D_{i, t}^{s}
\end{equation}
where $s = c$ for the chartist strategy and $s = f$ for the fundamentalist strategy. 

Finally, the i-th LF trader determines their probability of being chartist in the next session according to
\begin{equation}
\Phi_{i, t}^{c} = \frac{e^{\pi_{i, t} ^{c} / \zeta}}{e^{\pi_{i, t} ^{c} / \zeta} + e^{\pi_{i, t} ^{f} / \zeta}}
\end{equation}
where $\zeta > 0$ is the intensity of switching parameter.

\subsection{HF Trader Agent Specification}

Each simulation contains $N_H$ HF traders, where different HF traders are represented by the index $j = 1,...,N_H$. In contrast to LF traders, HF traders in the model follow an event-based activation procedure as opposed to a chronological trading frequency. This is again consistent with the work of \citet{Easley12}, in which it is suggested that HF traders follow an event time paradigm.

In an arbitrary session, $t \leq T$, the j-th HF trader will activate should the following activation condition be satisfied
\begin{equation}
\left| \frac{\bar{P}_{t - 1} - \bar{P}_{t - 2}}{\bar{P}_{t - 2}} \right| > \Delta x_{j}
\end{equation}
where $\Delta x_{j}$ is drawn from a truncated uniform distribution, with support between $\eta_{min}$ and $\eta_{max}$. 

Following activation, active HF traders either buy or sell, where this choice is made randomly with equal probability. In a construction again consistent with the empirical findings of \citet{Easley12}, HF trader agents in the model exploit order book dynamics created by LF trader agents earlier in each trading session by setting their order sizes according to order volumes present in the order book, in an attempt to absorb the orders of LF traders. This is represented by each active HF trader in the simulation setting their order size randomly according to an exponential distribution \footnote{In the original paper by \citet{JacobLeal15}, a Poisson distribution is used, where the mean volume is weighted by $0<\lambda<1$. We retain this weighting by $\lambda$, but discussions on why the distribution was changed can be found in Section \ref{Model}.\ref{Modifications}.}, with the mean being the average order volume in the opposite side of the order book. For an HF trader who is selling units of the asset, the opposite side of the order book would consist of buy orders and vice-versa.

After the setting of the order size as previously discussed, the j-th HF trader sets their order price (provided they are active) according to
\begin{equation}
P_{j, t} = P_t^{bid}(1 - \kappa_J) \label{HFOrderPriceA}
\end{equation}
if selling, or according to
\begin{equation}
P_{j, t} = P_{t}^{ask}(1 + \kappa_j) \label{HFOrderPriceB}
\end{equation}
if buying, where $P_t^{bid}$ and $P_t^{ask}$ are the best bid and ask in the order book respectively for session $t$ and $\kappa_j$ is a uniform random variable with support between $\kappa_{min}$ and $\kappa_{max}$.

Once the order is placed, it remains in the LOB until it is matched or until $\gamma^H < \gamma^L$ sessions have passed, resulting in its removal from the order book. Finally, at the end of each session, the j-th HF trader calculates their profit according to: 
\begin{equation}
\pi_{j, t} = (\bar{P}_{t} - P_{j, t})D_{j, t}
\end{equation}
where $D_{j, t}$ is the j-th HF trader's order size for the session.

\subsection{Matching Engine and Market Clearing \label{Matching}}

\citet{JacobLeal15} indicate that orders are to be matched by price and then time priority. In our implementation, it is noted that after all trader agents have placed orders in a particular session, a crossing of the bid-ask spread is typically observed. In other words, $P_t^{ask} - P_t^{bid} < 0$ after all orders have been placed. This implies that the best bid and best ask can immediately be matched, resulting in a trade. 

\begin{figure}[H]
\centerline{
\includegraphics[width=0.9\linewidth]{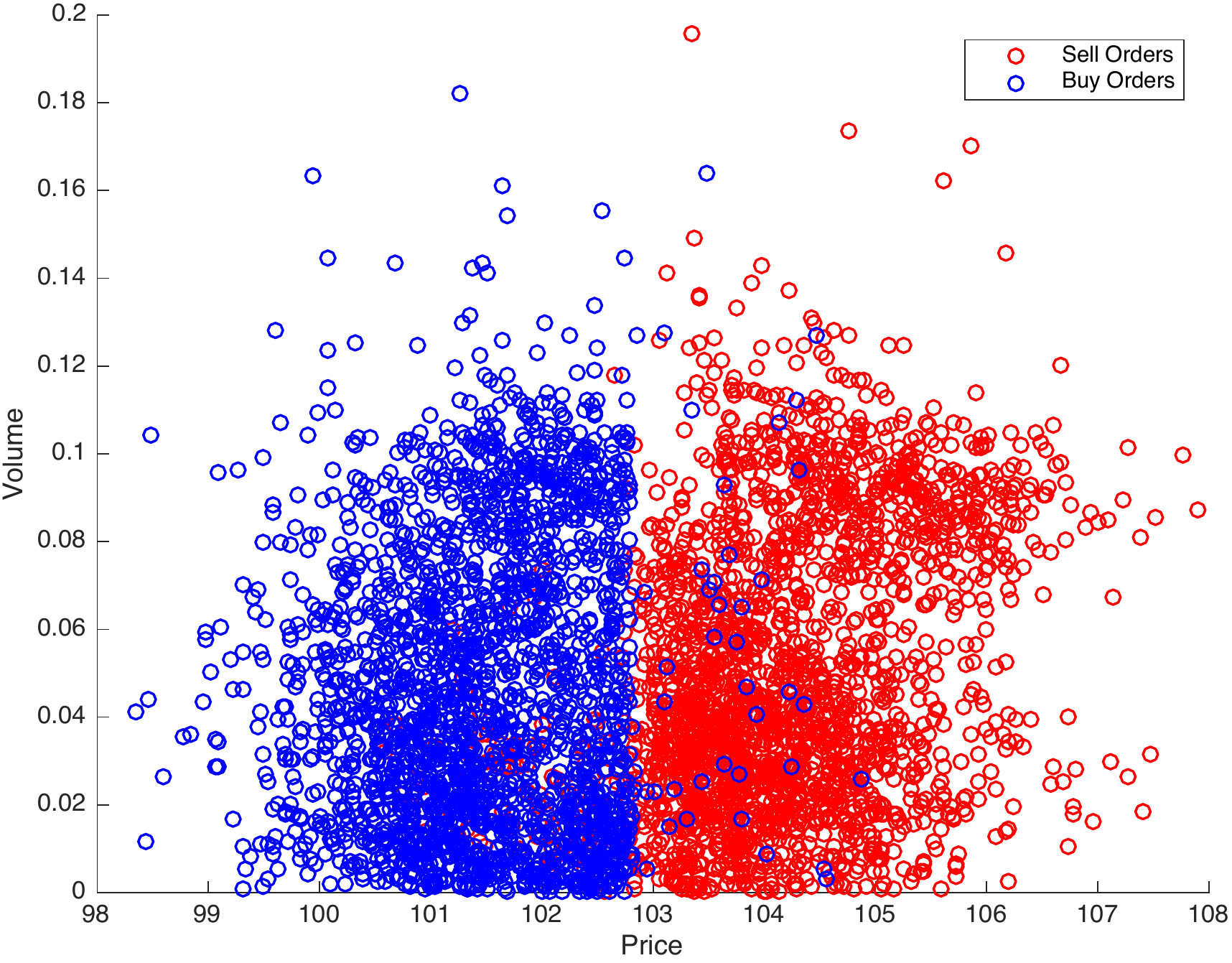}
}

\centerline{
\includegraphics[width=0.9\linewidth]{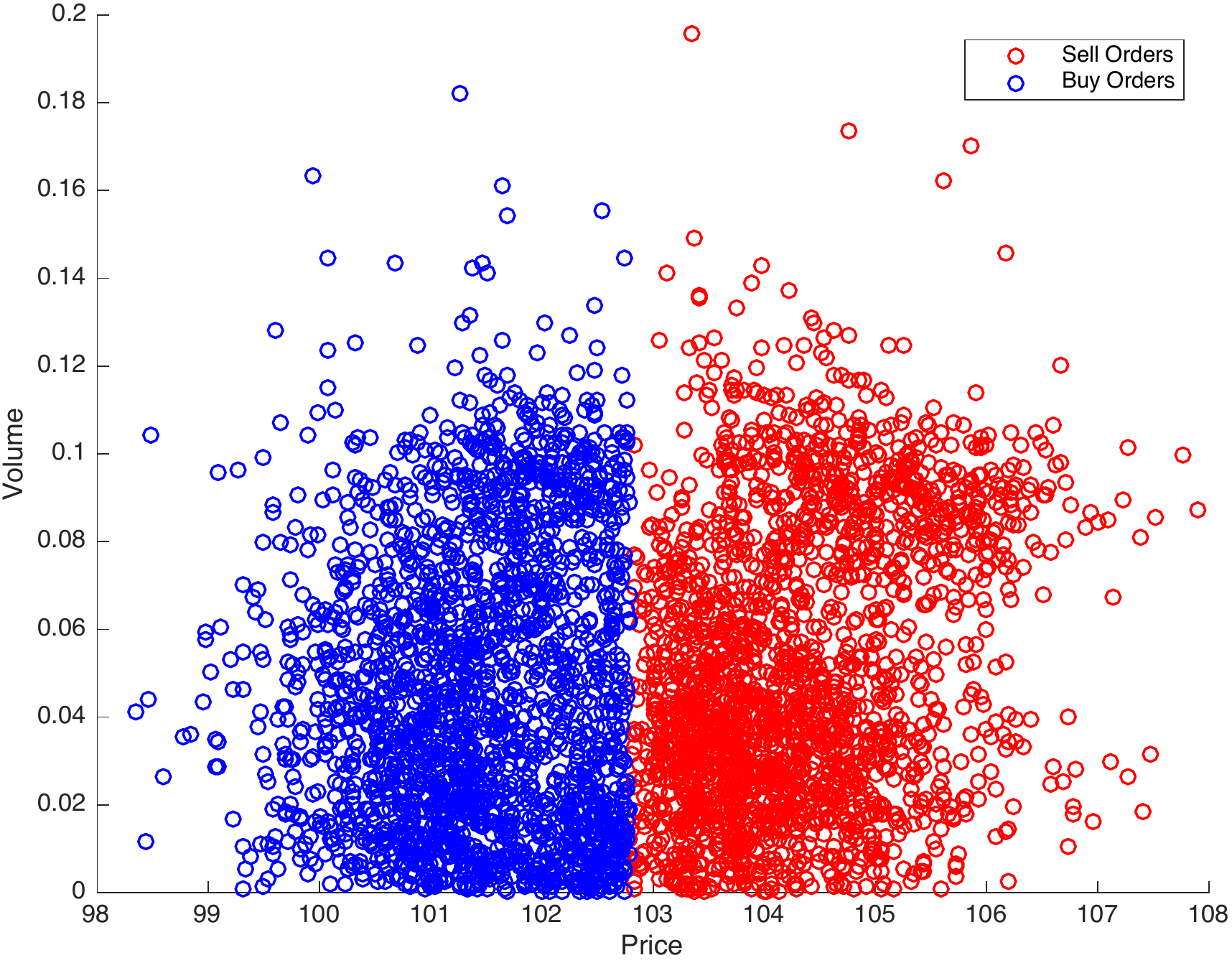}
}
\caption{Illustration of a typical limit order book configuration before (top) and after (bottom) order matching for an arbitrary session, showcasing an initial crossing of the bid-ask spread}
\end{figure}

We set the trade price equal to the average of the prices of the matched buy and sell orders and the trade size equal to the size of the smallest order in the matched pair. We can thus execute all executable limit orders at the end of each session by continually matching the best bid to the best ask until the bid-ask spread is no longer crossed, with the best bid and best ask changing as orders are depleted. Should there be a tie for the best bid or best ask, the order that was placed earliest is always executed first. The market price for each session is simply the final trade price for that session.

It is evident that the model's mechanism of batch order placement by all active traders before matching orders at the end of each session is not realistic. In a true, continuous double auction market, attempts to match orders would be made continuously, as orders are placed, rather than at the end of a specified period. Markets of this type would also involve additional order types, such as market orders. 

The order matching processes present in the model are therefore market clearing processes reminiscent of closing auctions at the end of a trading day rather than the continuous matching found in double auction markets during intraday periods. This is, however, fairly standard practice in many other models, notably that of \citet{Preis06}.

\subsection{Required Modifications \label{Modifications}}

We implement the model as it was originally presented, with two minor changes. 

Firstly, we implement our own method of initialization. This stems from the fact that model initialization procedures are not discussed in detail by \citet{JacobLeal15}. In order to be consistent with the opening auctions found at the start of each trading day in actual double auction markets, we begin with an analogous event in the first simulated trading session. In this session, all LF traders in the simulation place limit orders to buy or sell according to a random initial strategy, where the probability of each LF trader being either chartist or fundamentalist is initially equal. This mimics the large spike in activity that can be observed at the start of a trading day, an aspect of intraday seasonality \citep{Lehalle13, Cartea15}.

Secondly, for the parameters originally suggested by \citet{JacobLeal15}, $\alpha^c = 0.04$, $\sigma^c = 0.05$, $\alpha^f = 0.04$ and $\sigma^f = 0.01$, as well as an initial market price of $\bar{P}_0 = 100$ and an initial fundamental value of $F_0 = 100$, it was found that the model produced LF trader order sizes typically less than 1 under both the chartist and fundamentalist strategies. Referring to Eqn. (\ref{Chartist}), it is evident that even for relatively large price changes over a one-minute period ($\sim10$ units of currency), the model will produce order sizes typically less than 1 for the given parameters. Even if $\alpha^c$ and $\sigma^c$ were to be made larger, the fact that $\alpha^c$ is bounded above by 1 means that typical price changes ($\sim1$ units of currency) will still result in the same problematic order size dynamics. Analogous issues apply to Eqn. (\ref{Fundamentalist}). This in itself is somewhat problematic, as real markets only allow for whole number order sizes. 

Furthermore, the fact that HF trader order sizes were originally set according to a Poisson distribution, with mean dependent on the average order size on the opposite side of the order book, resulted in the majority of active HF trader order sizes being 0 when combined with these unusual LF trader order sizes. For this reason, an exponential distribution was used instead, allowing for the unusually small mean order sizes found in the model to remain as is without resulting in 0 values for HF trader order sizes. Exponential distributions are used to determine order sizes in many models throughout existing literature, a notable example being given by \citet{Mandes15}. This provides suitable justification for this choice of distribution. 

It should be noted that when matching orders, the only consideration with regards to order sizes is that they be sensible in relation to one another, with actual magnitudes being of more importance for studies involving trade volume effects. Thus, for our purposes, this simple correction is sufficient and these unusual order sizes will not affect the dynamics of the obtained market price time series. A more comprehensive correction of this order size issue should be considered in studies in which trade volumes and order sizes are important characteristics.

\section{Replication of Stylized Facts \label{StylizedFacts}}

Considering the minor changes made, it is essential that we demonstrate the ability of our implementation of the model to reproduce well-known stylized facts of return time series\footnote{A study of stylized facts in the context of the JSE is presented by \citet{Wilcox08}.} and provide some evidence of the similarity between our results and those obtained by \citet{JacobLeal15}.

\begin{table}[H]
\caption{Parameters for Stylized Fact Replication \label{JLParameters}}
\begin{tabular}{cccc}
\hline
Parameter & Value & Parameter & Value \\ \hline
T & 1200 & $\sigma_y$ & 0.01 \\ 
$N_L$ & 10000 & $\delta$ & 0.0001 \\
$N_H$ & 100 & $\sigma^z$ & 0.01 \\
$\theta$ & 20 & $\zeta$ & 1 \\
$[\theta_{min}, \theta_{max}]$ & $[10, 40]$ & $\gamma^L$ & 20 \\
$\alpha^c$ & 0.04 & $\gamma^H$ & 1 \\
$\sigma^c$ & 0.05 & $[\eta_{min}, \eta_{max}]$ & $[0, 0.2]$ \\
$\alpha^f$ & 0.04 & $\lambda$ & 0.625 \\
$\sigma^f$ & 0.01 & $[\kappa_{min}, \kappa_{max}]$ & $[0, 0.01]$ \\ \hline
\end{tabular}
\vspace{0.2cm}
\caption*{Parameters for simulations used to verify the implemented model's ability to replicate well-known return time series stylized facts, set according to those specified in Table 4 in the work conducted by \citet{JacobLeal15}}
\end{table}

\vspace{-0.5cm}

We provide this required evidence by simulating series of prices using the implemented model and using these prices to calculate series of log returns. We generate a total of $50$ price paths (Monte Carlo replications) using the default parameters provided by \citet{JacobLeal15}. We then proceed to construct a histogram approximating the distribution of the obtained log returns pooled over all the conducted Monte Carlo replications, along with a fitted normal density function, as well as a Q-Q plot for the same simulated data. Thereafter, we consider the autocorrelation functions of the returns and absolute values of the returns and provide the associated $95\%$ confidence bands.

\begin{figure}[H]
\centerline{
\includegraphics[width=0.9\linewidth]{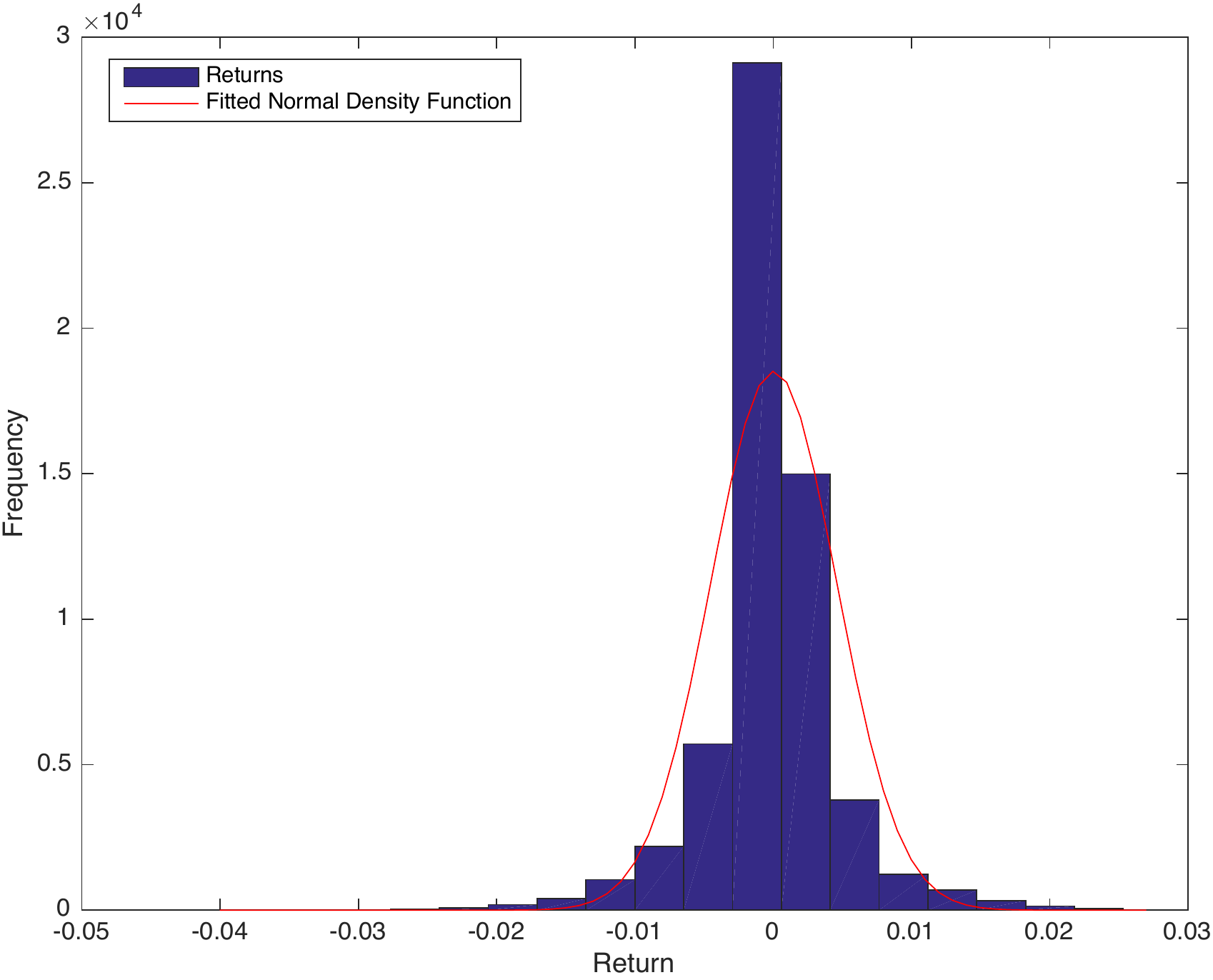}
}

\caption{Histogram representing the distribution of the simulated log returns \label{Returns}}
\end{figure}

\begin{figure}[H]
\centerline{
\includegraphics[width=0.9\linewidth]{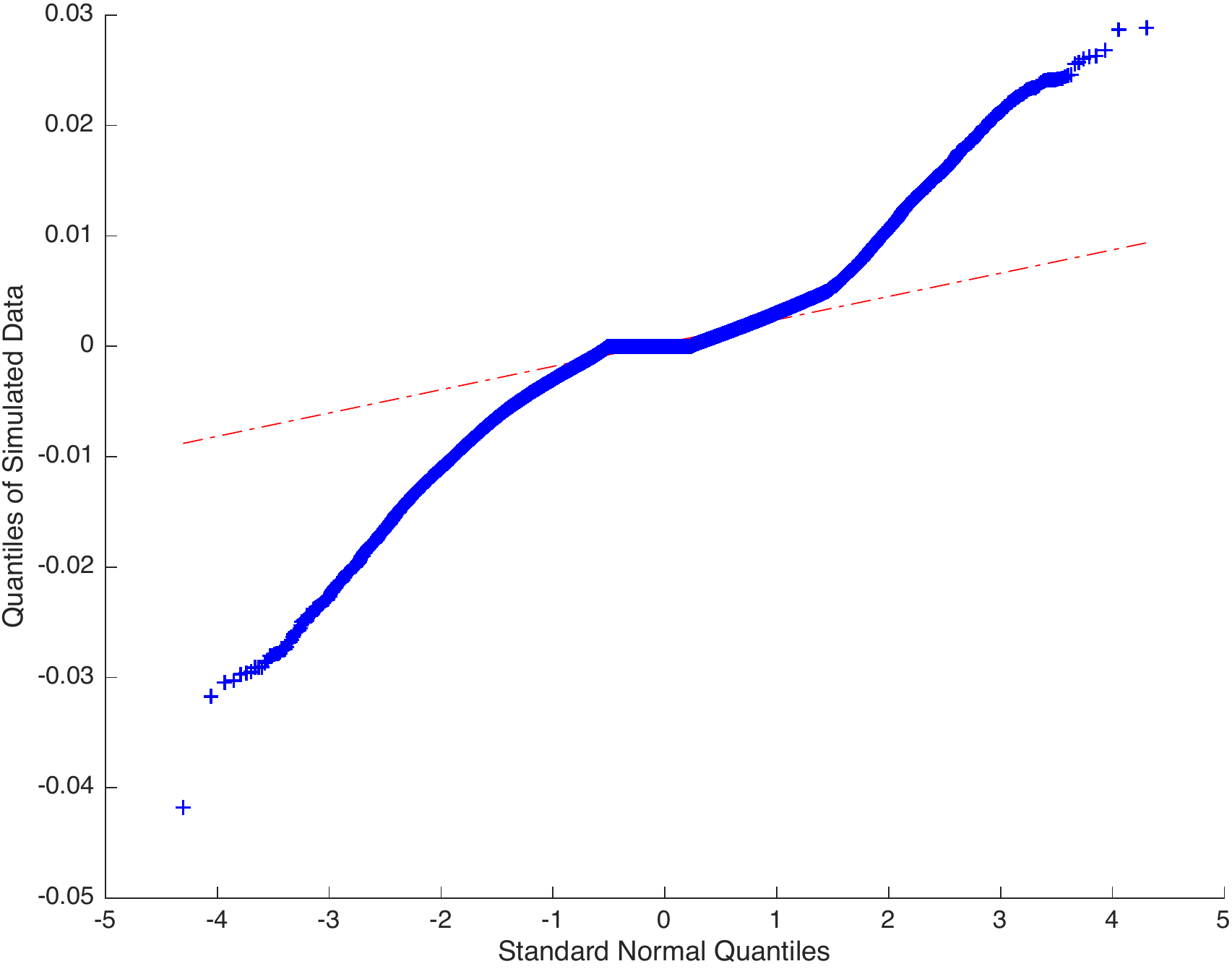}
}

\caption{Q-Q plot comparing the distribution of the simulated log returns to a normal distribution \label{QQPlot}}
\end{figure}

As is evident in Figures \ref{Returns} and \ref{QQPlot}, the distribution of the simulated log returns is leptokurtic and fat-tailed when compared to a normal distribution.

\begin{figure}[H]
\centerline{
\includegraphics[width=0.9\linewidth]{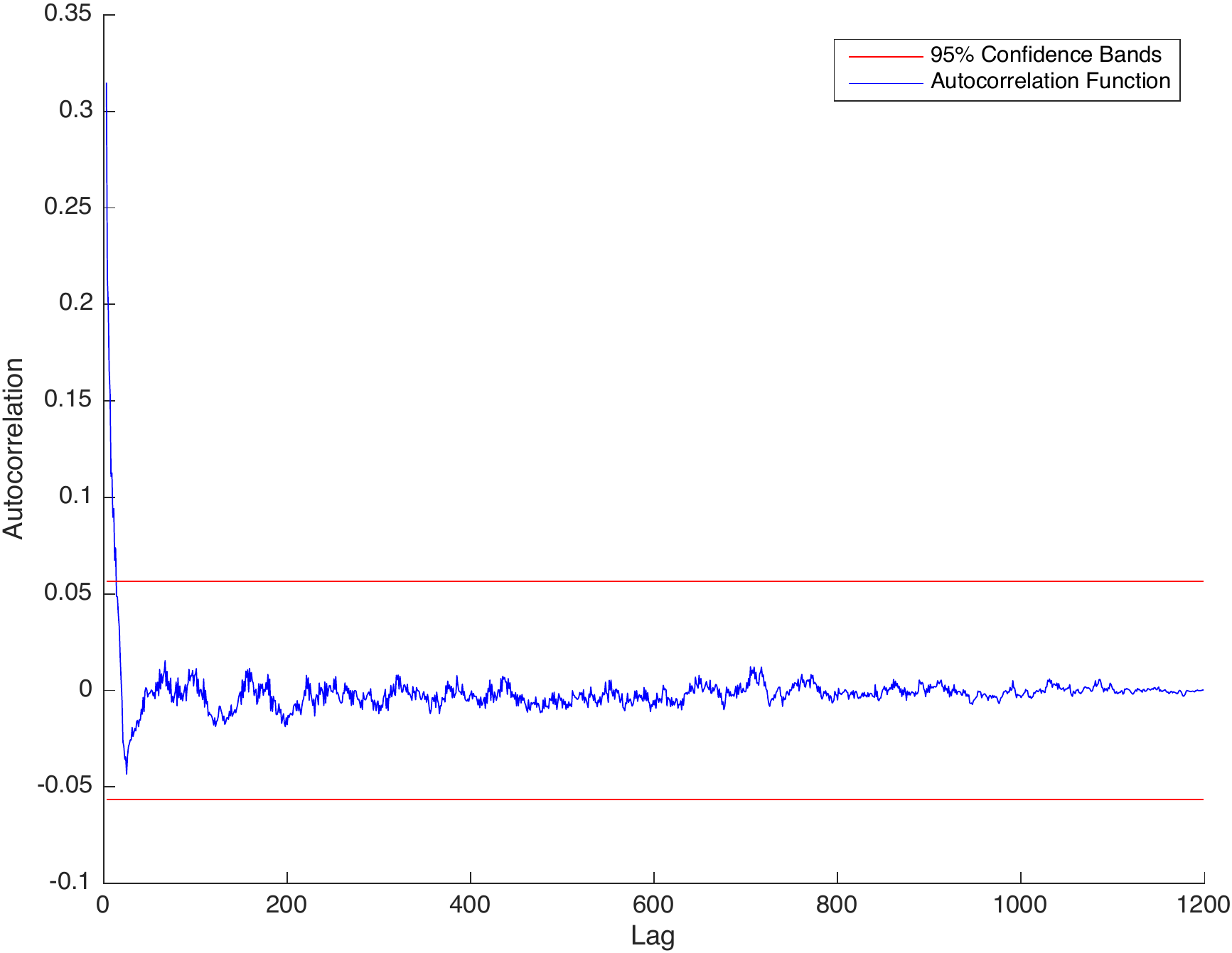}
}

\caption{Autocorrelation function for the series of log returns, with one-minute session lags \label{Autocorr}}
\end{figure}

\begin{figure}[H]
\centerline{
\includegraphics[width=0.9\linewidth]{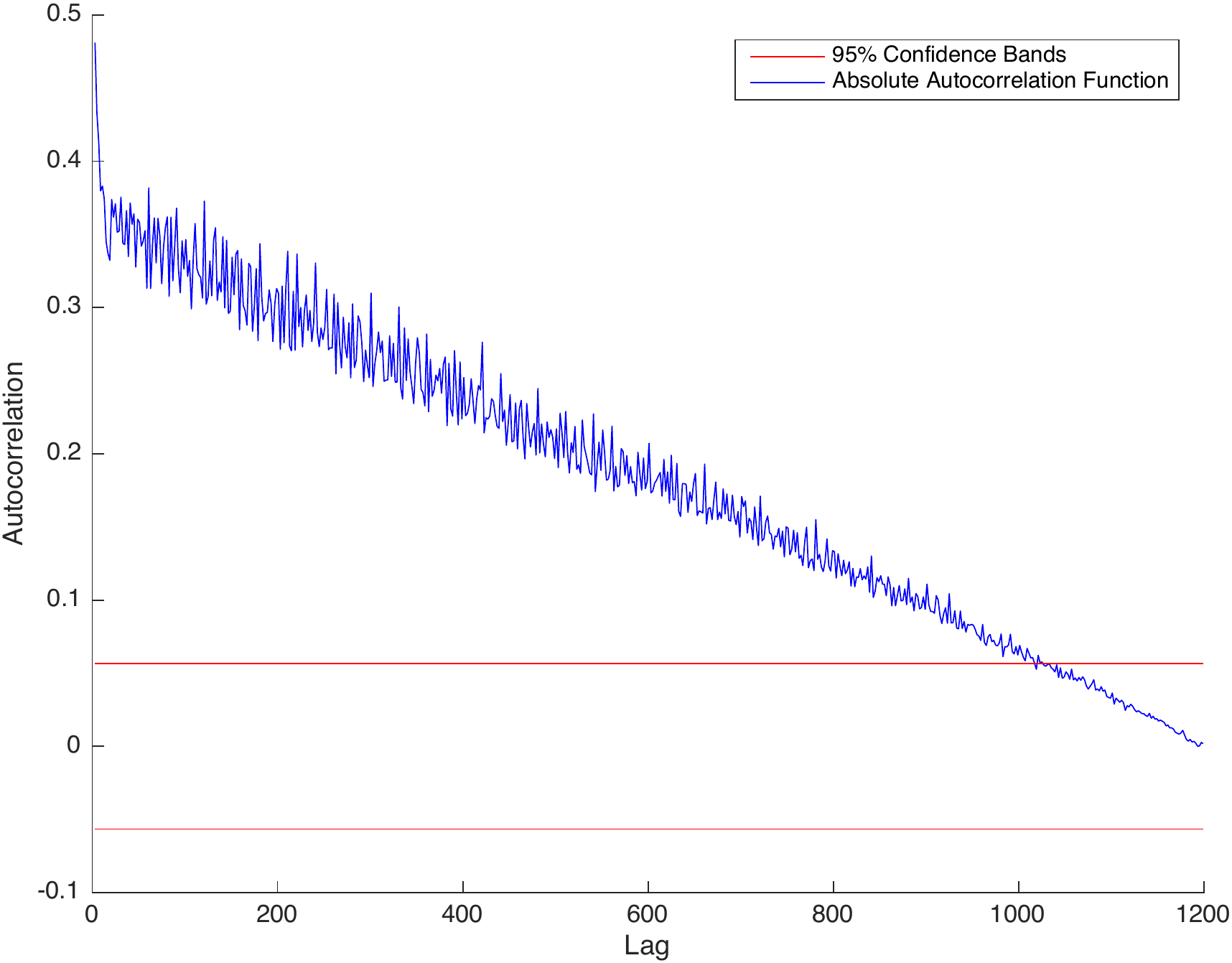}
}

\caption{Autocorrelation function for the series of the absolute values of the log returns, with one-minute session lags \label{Absolute_Autocorr}}
\end{figure}

Referring to Figure \ref{Autocorr}, the series of simulated log returns demonstrates behaviors consistent with those presented in the empirical investigations of \citet{Cont01}, with no significant pattern or deviation from zero being observed in the autocorrelation function for almost all lags. When considering Figure \ref{Absolute_Autocorr}, however, we observe initially significant autocorrelation for shorter lags with a steady decline in autocorrelation as the lag increases in the case of the series of the absolute values of the simulated log returns. The latter of these observations can be taken as evidence of volatility clustering \citep{Cont97}. 

All of the aforementioned properties are presented as key stylized facts of empirically-observed return time series by \citet{Cont01}. It is therefore apparent that our implementation of the model is able to reproduce well-known log return times series stylized facts and generates behaviors consistent with those demonstrated by \citet{JacobLeal15}.

It is worth noting that \citet{JacobLeal15} consider only the stylized facts of log return time series in the process of model validation. In general, the stylized facts of the central LOB have become increasingly important in recent years and the exclusion of these characteristics is problematic if relying on stylized facts for model validation. A comprehensive survey of LOB stylized facts is provided by \citet{Bouchaud02}. It may also be worth considering the Hurst exponent, as has been done by \citet{Preis06}.

\section{Calibration Framework}

We consider the approach presented by \citet{Fabretti13}. Considering the framework's success in existing calibration experiments involving the \citet{Farmer02} model, we reproduce it without alteration where possible in an attempt to ascertain if the framework produces equally successful results when applied to intraday models. 

The calibration problem is formally defined as follows:

\begin{equation}
\min_{\theta \in \Theta} f(\theta) \label{Optimization}
\end{equation}
where $\theta$ is a vector of parameters, $\Theta$ is the space of feasible parameters and $f$ is the objective function.

\subsection{Objective Function Construction}

We construct the objective function for the considered dataset using the method proposed by \citet{Winker07}.

The transaction data set is obtained in the TRTH \citep{Reuters16} format, presenting a series of trades, quotes and auction quotes. We convert the data set to a series of one-minute price bars, with each price corresponding to the final quote mid price for each minute, where the mid price of a quote is given as the average of the level 1 bid price and level 1 ask price associated with that quote. From this series of prices, we may obtain a series of log prices \footnote{While \citet{Fabretti13} chooses log prices for convenience, it was found that the weight matrix required by the method was extremely ill-conditioned when considering either prices or log returns, motivating our choice to retain this convention.}, which is the series we attempt to calibrate the model to.

We make use of a combination of $k = 5$ moments and test statistics to represent the statistical properties of the log price time series previously mentioned, remaining consistent with \citet{Fabretti13}. We use:
\begin{enumerate}
\item The mean, representing the central tendency of the measured series.
\item The standard deviation, giving an indication of the spread of the measured series.
\item The kurtosis, acting as a measure of the mass contained in the tails of the probability distribution of the measured series.
\item The Kolmogorov-Smirnov (KS) test, comparing the empirical CDFs of the log price time series obtained from the transaction data and another simulated using the model.
\item The generalized Hurst exponent \citep{DiMatteo07}, with $q = 1$, representing the scaling properties of the measured series.
\end{enumerate}
We define the:
\begin{enumerate}
\item \textit{realized log prices}, $\mathbf{R} = \{r_t\}, t = 1,...,T$, to be a realization of the log price process, drawn from transaction data.
\item \textit{exact moments}, $\mathbf{m} = (m_1,...,m_k)$, to be the true values of the $k$ selected moments of the log price process. We cannot precisely know these true values; we can only estimate their values according to a particular realization of the log price process.
\item \textit{estimated moments}, $\mathbf{m}^e = (m_1^e,...,m_k^e)$, to be the estimate of $\mathbf{m}$ based on $\mathbf{R}$.
\item \textit{simulated moments}, $\mathbf{m}_i^s(\theta)$, to be the estimate of the $k$ selected moments for the simulated series, generated using parameter set $\theta$, where the index $i$ corresponds to the estimate associated with a particular simulation run.
\end{enumerate}

Using the method of simulated moments, we require:
\begin{equation}
\mathbb{E}[(\mathbf{m}^s|\theta)]=\mathbf{m}
\end{equation}
which yields:
\begin{equation}
\mathbb{E}[(\mathbf{m}^s|\theta) - \mathbf{m}]=0
\end{equation}
Assuming $\mathbf{m}$ is given and calculating the required expectation as the arithmetic average:
\begin{equation}
\frac{1}{I} \sum_{i = 1} ^{I}[(\mathbf{m}_i^s|\theta) - \mathbf{m}] = 0
\end{equation}
Using our estimate, $\mathbf{m}^e$, we set:
\begin{equation}
\hat{G}(\theta) = \frac{1}{I} \sum_{i=1}^{I}[(\mathbf{m}^s_i|\theta) - \mathbf{m}^e]
\end{equation}

For most of our experiments, we consider $I=5$, with any deviations from this convention stated. This limited number of Monte Carlo replications is selected primarily due to computational constraints, since a single simulation is capable of taking a fairly significant amount of time to complete. 

Roughly $7$ hours of computational time is required by each Nelder-Mead simplex experiment and roughly $10$ hours of computational time is required by each of the genetic algorithm experiments, despite being parallelized and run on a relatively powerful desktop computer \footnote{HP Z840 workstation with $2$ Intel Xeon E5-2683 CPUs, providing $2\times16=32$ workers.}.

While this is not ideal, computational difficulties remain a major obstacle in ABM calibration \citep{Grazzini15b} and are difficult to solve in practice, especially when simulations are complex and involve realistic matching processes. Nevertheless, it is important to note this relevant caveat.

Our objective function is finally given by:
\begin{equation}
f(\theta) = \hat{G}(\theta)^T \mathbf{W} \hat{G}(\theta)
\end{equation}
where $\mathbf{W}$ is a $k \times k$ matrix of weights.

$\mathbf{W}$ is set to be the inverse of the variance-covariance matrix of $\mathbf{m}^e$, denoted $Cov[\mathbf{m}^e$]. This selected weight matrix takes the uncertainty of estimation associated with $\mathbf{m}^e$ into account and assigns larger weights to moments associated with greater uncertainty and vice-versa. 

We can estimate $Cov[\mathbf{m}^e$] by applying a moving block bootstrap \citep{Kunsch89} to $\mathbf{R}$ with a window length of $b$ to create $n$ bootstrapped samples and then use these bootstrapped samples to calculate approximate distributions for each estimated moment or test statistic. Thereafter, we can use these distributions to calculate a covariance matrix \footnote{Specifically, we set $Cov[\mathbf{m}^e]_{i, j} = Cov[m_i^{e}, m_j^{e}]$, where $Cov[a, b]$ corresponds to the empirical covariance of $a$ and $b$.} approximating $Cov[\mathbf{m}^e$].

For all data sets considered, we set $b = 100$ and $n = 10000$, again consistent with \citet{Fabretti13}.

\subsection{Heuristic Optimization Methods}

The heuristic optimization methods employed in the minimization of the aforementioned objective function, namely the Nelder-Mead simplex algorithm with threshold accepting \citep{Gilli03} and genetic algorithm \citep{Holland75}, are presented in Appendix \ref{SimplexDescription} and Appendix \ref{GADescription} respectively.

\section{Data}

The considered dataset is acquired in the TRTH \citep{Reuters16} format, presenting a tick-by-tick series of trades, quotes and auction quotes. 

As previously stated, we convert the dataset to a series of one-minute price bars, with each price corresponding to the final quote mid price for each minute, where the mid price of a quote is given as the average of the level 1 bid price and level 1 ask price associated with that quote. From this series of prices, we obtain a series of log prices, which is the series we attempt to calibrate the model to. The rationale behind this sampling strategy is as follows: 

Trades appear more infrequently in the transaction data than in the model, where trades occur every minute. For this reason, the sampling of trade prices becomes problematic, with a large number of the one-minute periods considered in the data containing no trades. Quotes were therefore chosen instead, as we typically have at least one quote available during each minute of trading in the transaction data. 

The consideration of quote mid prices is an attempt to be consistent with the fact that trade prices in the \citet{JacobLeal15} model are essentially also mid prices and the final quote mid price was considered as the market price for each minute since the final trade price was selected as the market price in the model. 

It is evident, however, that reconciling an event-based intraday transaction dataset and a model that produces a series of chronologically-sampled prices is by no means intuitive and for this reason the above assumptions were necessary.

The transaction dataset often presents events occurring outside of standard trading hours, 9:00 to 17:00, but we consider only quotes occurring in the period from 9:10 to 16:50 on any particular trading day. This is necessary as the opening auction occurring from 8:30 to 9:00 tends to produce erroneous data during the first 10 minutes of continuous trading and the fact that the period from 16:50 to 17:00 represents a closing auction. 

In all calibration experiments, we consider one-week periods, corresponding to a total of 2300 one-minute price bars, representing 460 minutes of trading each day from Monday to Friday. We consider a single stock, Anglo American PLC, over the period beginning at 9:10 on 1 November 2013 and ending at 16:50 on 5 November 2013. The associated weight matrix required by the method of simulated moments is presented in Appendix \ref{WeightMatrix}.

\section{Calibration Results \label{CalResults}}

\subsection{Free and Fixed Parameters}

For most of our calibration experiments, we consider 10 free parameters, with the remainder of the parameters set to be fixed. More specifically, we set $\alpha^c$, $\alpha^f$, $\sigma^c$, $\sigma^f$, $\sigma^y$, $\delta$, $\sigma^z$, $\lambda$, $N_L$ and $N_H$, defined in Section \ref{Model}, to be free parameters and randomly generate initial values for these parameters during each calibration experiment. We initialize $N_H$ and $N_L$ between $100$ and $10000$ for each population member or simplex vertex and initialize the remaining free parameters with values between $0$ and $0.1$. This particular set of free parameters was selected due to the fact that it represents a broad range of model characteristics, including order size, order price and trader population dynamics, while still remaining computationally tractable. The results related to this free parameter set are presented in Sections \ref{CalResults}.\ref{NMSection} to \ref{CalResults}.\ref{ParamAnalysis}.

It must be noted that in the above, as is the case in the work of \citet{Fabretti13}, we consider more free parameters ($10$) than objective function moments ($5$). This particular feature of the investigation may be open to criticism and must be acknowledged and dealt with to ensure the validity and robustness of our conclusions. We therefore also perform supplementary calibration experiments in which we consider a smaller free parameter set including only the following parameters: $\delta$, $\sigma^z$, $N_H$ and $N_L$. In other words, we consider fewer free parameters than objective function moments and demonstrate that the results obtained in these experiments are indeed consistent with those obtained when considering more free parameters than objective function moments. The results related to this free parameter set are presented in Section \ref{CalResults}.\ref{Supplementary}.

In both of the above cases, all other parameters are constant and their values are set to be identical to those presented in Table \ref{JLParameters}, with $\bar{P}_0 = F_0$ and $\bar{P}_1$ set to be the first two prices of the measured price time series from which we calculate the series of measured log prices.

It is important to ensure that the initial value to which the simulated price time series is set does not happen to be an outlier of the empirically-observed price time series. In order to demonstrate this, we apply the well-known boxplot method proposed by \citet{Tukey77}, in which we are required to determine the interval 
\begin{center}
$[Q_1 - 1.5 IQR, Q_3 + 1.5 IQR]$, 
\end{center}
where $Q_1$ is the lower quartile, $Q_3$ is the upper quartile and $IQR$ is the interquartile range of the measured price time series.

Applying this method to our processed data set, we obtain the interval
\begin{center}
$[217.2638, 278.3337]$,
\end{center}
with any points outside of this interval predicted to be potential outliers. Therefore, our chosen value of $\bar{P}_0 = F_0 = 238.75$ is well within the interval and we can therefore say with relative confidence that our chosen initial price is not an outlier.

\subsection{Nelder-Mead Simplex Algorithm \label{NMSection}}

Due to the fact that we consider $n = 10$ free parameters in our calibration experiments, we begin with $n + 1 = 11$ simplex vertices, each consisting of randomly generated initial values for the $10$ free parameters. 

In our experiments, it was noted that $250$ iterations of the Nelder-Mead simplex algorithm with threshold accepting almost always produced convergent behavior and we thus conducted $20$ calibration experiments with this fixed number of iterations. Despite the presence of convergent behavior, the obtained parameter sets were surprisingly different for various calibration experiments, with a significant dependence on the set of initial vertices. This is strongly indicative of convergence to local minima, even with the inclusion of methods aiming to overcome this problem, namely the threshold accepting heuristic. 

More optimistically, however, a fairly significant number of these experiments (more than half) seemed to converge to parameter sets with objective function values in a relatively small range, a similar result to that obtained by \citet{Fabretti13}, indicating that our implementation has, at the very least, produced similar behaviors when attempting to calibrate the model. Furthermore, the obtained parameter sets in this region of similar objective function values produced model behaviors that were fairly consistent with those demonstrated by the empirical data, with the error between the simulated and measured moments, notably the mean, standard deviation and Hurst exponent, being within similar ranges to those obtained by \citet{Fabretti13}. An assessment of the quality of the model fit is presented in Section \ref{CalResults}.\ref{DataComparison}.

\begin{table}[H]
\caption{Nelder-Mead Simplex Calibration Results}
\begin{tabular}{ccc}
\hline
Parameter & 95\% Conf Int & $\frac{s}{\sqrt{n}}$ \\ \hline
$\alpha^c$ & $[0.0249, 0.111]$ & $0.0205$ \\
$\sigma^c$ & $[0.0322, 0.0939]$ & $0.0147$ \\
$\alpha^f$ & $[0.0299, 0.0597]$ & $0.00711$ \\
$\sigma^f$ & $[0.0389, 0.0747]$ & $0.00853$ \\
$\sigma^y$ & $[0.00689, 0.0547]$ & $0.0114$ \\
$\delta$ & $[0.000391, 0.00212]$ & $0.000413$ \\
$\sigma^z$ & $[0.0159, 0.0429]$ & $0.00643$ \\
$\lambda$ & $[0.0250, 0.0696]$ & $0.0107$ \\
$N_L$ & $[2420, 5693]$ & $782.0283$ \\
$N_H$ & $[3611, 7746]$ & $987.594$ \\ \hline
\end{tabular}
\vspace{0.2cm}
\caption*{95\% confidence intervals for the set of free parameters, obtained from 20 independent calibration experiments involving the Nelder-Mead simplex algorithm combined with the threshold accepting heuristic \label{SimplexResults}}
\end{table}

\vspace{-0.5cm}

Referring to Table \ref{SimplexResults}, it is apparent that the calibration process, while able to reduce the search region from $[0, 0.1]$ and $[100, 10000]$ to the regions roughly specified by the confidence intervals \footnote{The intervals are calculated as: $\bar{x} \pm t^{*} \frac{s}{\sqrt{n}}$, where $\bar{x}$ is the sample mean, $s$ is the sample standard deviation, $n$ is the sample size, and $t^{*}$ is the appropriate critical value for the $t$ distribution.} shown, still produces results where the vast majority of these intervals are still far too large to identify any true convergence to a unique set of optimal parameters. Despite this general trend, however, we see an interesting irregularity. The parameter $\delta$ does indeed seem to demonstrate convergent behavior and the algorithm has identified a very small confidence interval for $\delta$. It is clear that this requires more thorough investigation and is discussed in detail in Section \ref{CalResults}.\ref{ParamAnalysis}.

Therefore, the obtained results are more or less indicative of an ability to find feasible parameter sets, but we find it problematic that we do not obtain stable or unique parameters and that there is evidence of a strong dependence on the initial simplex. It is evident that the method converges to reasonable solutions, but we cannot be sure that the obtained solutions truly minimize the objective function, which can be perceived as fairly problematic.

\subsection{Genetic Algorithm}

We found that convergence was observed in almost all of the genetic algorithm experiments after approximately $100$ generations. For this reason, we conducted $10$ calibration experiments, with random initial populations using the same initial parameter range as in the Nelder-Mead simplex calibration experiments, and iterated the populations for $100$ generations. This smaller number of calibration experiments was selected due to the increased computational cost of the genetic algorithm.

\begin{table}[H]
\caption{Genetic Algorithm Calibration Results}
\begin{tabular}{ccc}
\hline
Parameter & 95\% Conf Int & $\frac{s}{\sqrt{n}}$ \\ \hline
$\alpha^c$ & $[0.0367, 0.0773]$ & $0.00899$ \\
$\sigma^c$ & $[0.0459, 0.0840]$ & $0.00843$ \\
$\alpha^f$ & $[0.0234, 0.0706]$ & $0.0104$ \\
$\sigma^f$ & $[0.0187, 0.0708]$ & $0.0115$ \\
$\sigma^y$ & $[0.0332, 0.0932]$ & $0.0133$ \\
$\delta$ & $[0.000114, 0.00108]$ & $0.000213$ \\
$\sigma^z$ & $[-0.00222, 0.0144]$ & $0.00367$ \\
$\lambda$ & $[0.0468, 0.0924]$ & $0.0101$ \\
$N_L$ & $[469, 5337]$ & $1075.947$ \\
$N_H$ & $[3335, 7881]$ & $1004.866$ \\ \hline
\end{tabular}
\vspace{0.2cm}
\caption*{95\% confidence intervals for the set of free parameters, obtained from 10 independent calibration experiments involving the genetic algorithm \label{GAResults}}
\end{table}

\vspace{-0.5cm}

Referring to Table \ref{GAResults}, it is clear that the obtained calibration results point to similar conclusions as were obtained using the Nelder-Mead simplex algorithm. We obtain similar parameter confidence intervals for each calibrated parameter and again find that all parameters, with the exception of $\delta$, seem unable to be uniquely determined.

\subsection{Comparison of Calibrated Model and Empirical Data \label{DataComparison}}

It was mentioned in the preceding sections that while no unique parameter set could be established, a number of the experiments converged to different parameter sets with similar objective function values producing reasonable behavior when compared to the empirical data. In this section, we consider the best parameter sets, presented in Table \ref{BestParams}, obtained in both the Nelder-Mead simplex and genetic algorithm experiments, associated with Tables \ref{SimplexResults} and \ref{GAResults} respectively.

\begin{table}[H]
\caption{Best Parameter Sets Obtained Using Each Calibration Method \label{BestParams}}
\begin{tabular}{ccc}
\hline
Calibrated Parameter & $\theta_{NM + TA}$ & $\theta_{GA}$ \\ \hline
$\alpha^c$ & $0.0433$ & $0.0025$ \\
$\sigma^c$ & $0.0233$ & $0.0728$ \\
$\alpha^f$ & $0.0068$ & $0.0222$ \\
$\sigma^f$ & $0.0805$ & $0.0228$ \\
$\sigma^y$ & $0.0017$ & $0.0021$ \\
$\delta$ & $0.00004$ & $0.00002$ \\
$\sigma^z$ & $0.0309$ & $0.0238$ \\
$\lambda$ & $0.0657$ & $0.0149$ \\
$N_L$ & $9783$ & $9668$ \\
$N_H$ & $5150$ & $5949$ \\ \hline
\end{tabular}
\vspace{0.2cm}
\caption*{Best parameter sets obtained through the implementation of Nelder-Mead simplex algorithm combined with the threshold accepting heuristic and the genetic algorithm}
\end{table}

\vspace{-0.5cm}

For each of these parameter sets, we simulate $50$ price paths and obtain the $95\%$ confidence interval for the estimates of the mean, standard deviation, kurtosis, and Hurst exponent of the simulated log prices and compare these to the empirical moments obtained from the transaction data.

Referring to Table \ref{CalibratedMoments}, we see that the calibration strategy has resulted in log price paths with comparable means and standard deviations to that observed in the empirical data, along with a reasonable, but slightly underestimated generalized Hurst exponent estimate for the Nelder-Mead simplex parameters and a fairly accurate generalized Hurst exponent estimate for the genetic algorithm parameters. This indicates that the calibration experiments performed, though not conclusive in determining a unique parameter set, were able to produce reasonable fits for some of the moments. We also see, as was the case in the investigation of \citet{Fabretti13}, that the genetic algorithm tends to perform slightly better than the Nelder-Mead simplex algorithm with threshold accepting.

\begin{table}[H]
\caption{Comparison of Empirical and Calibrated Moments \label{CalibratedMoments}}
\begin{tabular}{cccc}
\hline
& $\theta_{NM + TA}$ & $\theta_{GA}$ & Data \\ \hline
$m_1$ & $[5.4931, 5.5436]$ & $[5.4450, 5.5273]$ & $5.5143$ \\
$m_2$ & $[0.0292, 0.0449]$ & $[0.0263, 0.0532]$ & $0.0300$ \\
$m_3$ & $[2.1092, 2.6760]$ & $[1.9029, 2.6317]$ & $1.2402$ \\
$m_4$ & $[0.5110, 0.5239]$ & $[0.5551, 0.5826]$ & $0.5658$\\ \hline
\end{tabular}
\vspace{0.2cm}
\caption*{$95\%$ confidence intervals for simulated moments obtained using calibrated parameters compared to the empirically measured moments, with $m_1,...,m_4$ corresponding to the mean, standard deviation, kurtosis and generalized Hurst exponent respectively}
\end{table}

\vspace{-0.5cm}

In contrast to this, the kurtosis is more significantly overestimated. This seems to stem from flaws in the calibration methodology. In both our own investigation and that of \citet{Fabretti13}, the kurtosis tended to be the most poorly fit moment after model calibration. This is likely explained by the fact that both our own weight matrix and that of \citet{Fabretti13} tends to assign a very small weight to errors on the kurtosis. Considering that the method employed was reproduced without alteration, it may be necessary for the choice of moments or the construction of the weight matrix to be revisited in future, since both our weight matrix and that of \citet{Fabretti13} tends to be significantly biased towards finding good fits for the mean and standard deviation at the expense of the other considered moments. A possible alternative to the currently constructed objective function is the information theoretic criterion proposed by \citet{Lamperti15} and applied in \citet{Lamperti16}.

This has, however, demonstrated that the method in question has indeed found reasonable fits for the most significantly weighted moments in the objective function.

\subsection{Parameter Analysis \label{ParamAnalysis}}

In an attempt to illustrate the behavior of the objective function with respect to changes in parameter values, we generate surfaces representing the objective function values corresponding to various parameter value pairs, as has been done by \citet{Gilli03} and \citet{Fabretti13}. The process of generating these surfaces is as follows:

\begin{enumerate}
\item We choose any pair of parameters drawn from the $10$ free parameters selected in the calibration experiments. 
\item We then define the space of possible values for each parameter in the pair. In general, we set the space to be $[0, 0.1]$ on $\mathbb{R}$ for all parameters, with the exception of $N_L$ and $N_H$, where we set the space to be $[100, 10000]$ on $\mathbb{R}$.
\item This results in a two-dimensional space whose x-axis is defined by the possible values of one of these selected parameters and whose y-axis is similarly defined by the possible values of the other parameter.
\item We then populate this two-dimensional space by generating $1000$ points using an appropriate two-dimensional Sobol sequence. In the case of $N_L$ and $N_H$, we round the values of the obtained points.
\item These randomly generated parameter values are then used to simulate $5$ price and hence log price time series (Monte Carlo replications) using the \citet{JacobLeal15} model, with all other free parameter values set according to $\theta_{NM + TA}$ in Table \ref{BestParams}.
\item We can thus evaluate the objective function at each of the generated points and obtain an approximate surface of objective function values corresponding to all possible values of the parameters in the pair using cubic interpolation.
\end{enumerate}

After the consideration of surfaces for a number of parameter pairs, we found that there are $3$ main types of surface behaviors that emerge, with only one of these behaviors resulting in satisfactory calibration results.

\begin{figure}[H]
\centerline{
\includegraphics[width=0.9\linewidth]{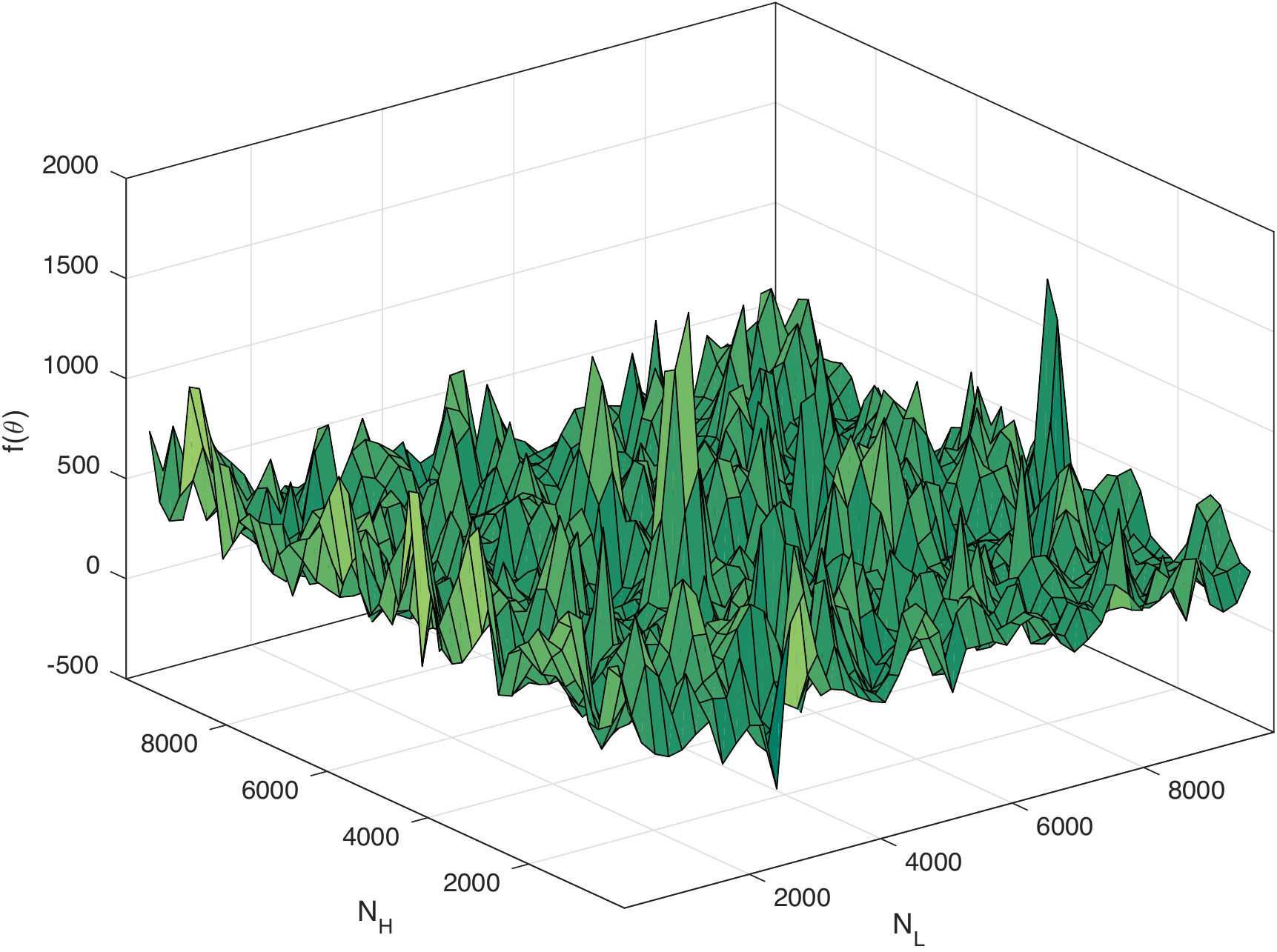}
}

\caption{Objective function surface demonstrating type 1 behavior: $N_H$ and $N_L$ \label{Surface_LFHFTraders}}
\end{figure}

Referring to Figure \ref{Surface_LFHFTraders}, illustrating what we will call type 1 behavior, we see that changes in $N_L$ and $N_H$ have a haphazard effect on the objective function, with very little evidence of structure. It is evident that even slight changes in these parameters have the potential to result in fairly large relative differences in objective function values, leading to a surface dominated by a series of peaks and valleys, with no clear path to a global minimum or discernable trend emerging. This is the likely reason that both the Nelder-Mead simplex algorithm with threshold accepting and genetic algorithm had a tendency to converge to local minima, even with attempts made in both search methods to avoid this. It seems very unlikely that the model can be calibrated in a meaningful way with respect to parameters demonstrating this type of behavior, as the observed degeneracies are fairly severe, with such parameters generating unpredictable, near-random effects on the considered objective function. Other parameters generating similar behaviors include $\alpha^c$, $\alpha^f$, $\sigma^c$, $\sigma^f$ and $\lambda$.

\begin{figure}[H]
\centerline{
\includegraphics[width=0.9\linewidth]{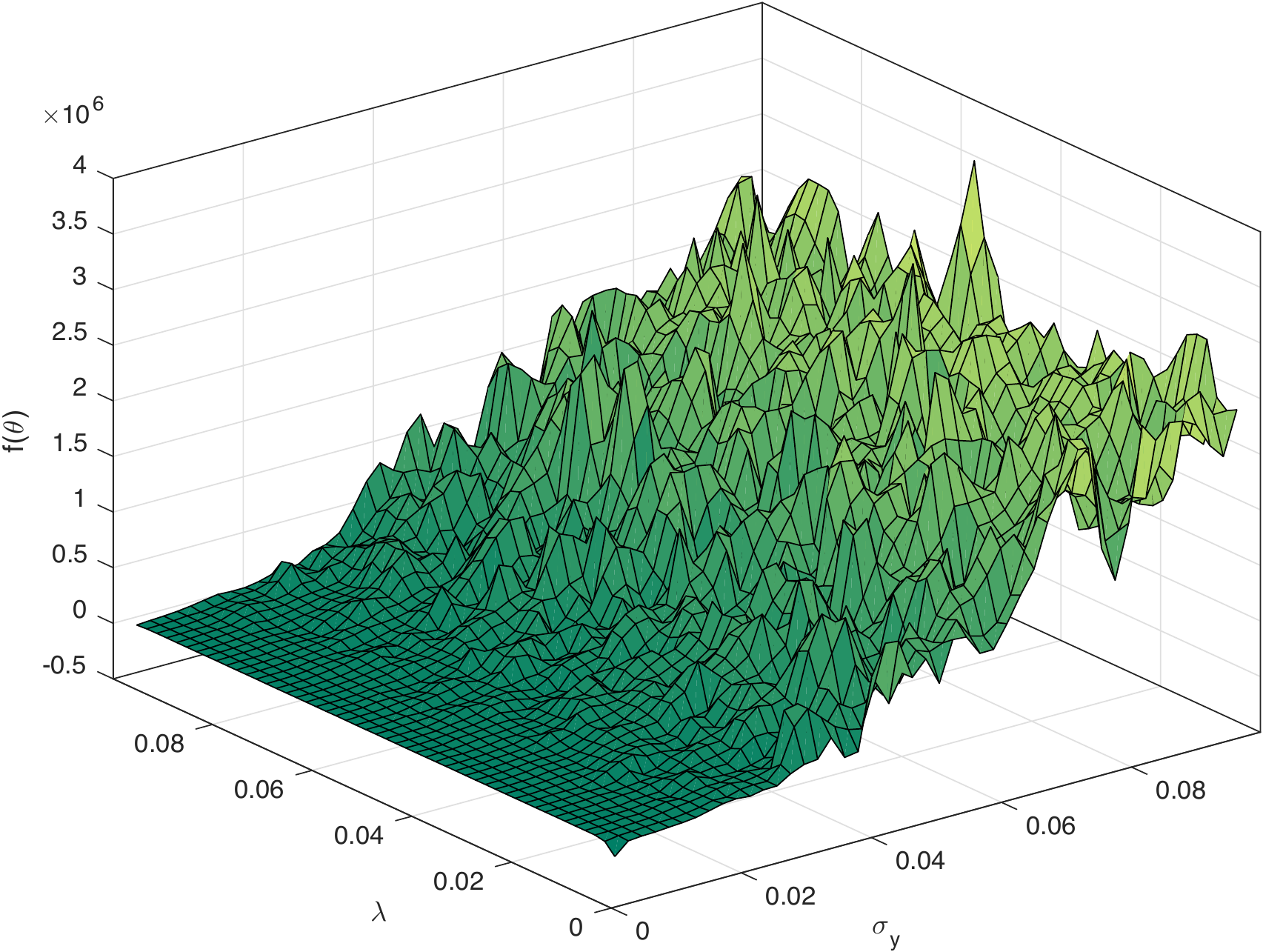}
}

\caption{Objective function surface demonstrating type 2 behavior: $\lambda$ and $\sigma^y$ \label{Surface_Lambda_Sigma_y}}
\end{figure}

Referring to Figure \ref{Surface_Lambda_Sigma_y}, illustrating what we call type 2 behavior, we see that there is indeed a trend presented by $\sigma^y$, albeit with many local minima and maxima obscuring this trend, resulting in a fairly noisy surface with some form of overall path to a global minimum emerging. While this type of behavior is indeed more promising than the first, both optimization methods failed to demonstrate any real convergence with respect to $\sigma^y$. This is most likely due to the presence of a very large number of local minima and a lack of smoothness hiding the observed trend from the optimization methods considered.

\begin{figure}[H]
\centerline{
\includegraphics[width=0.9\linewidth]{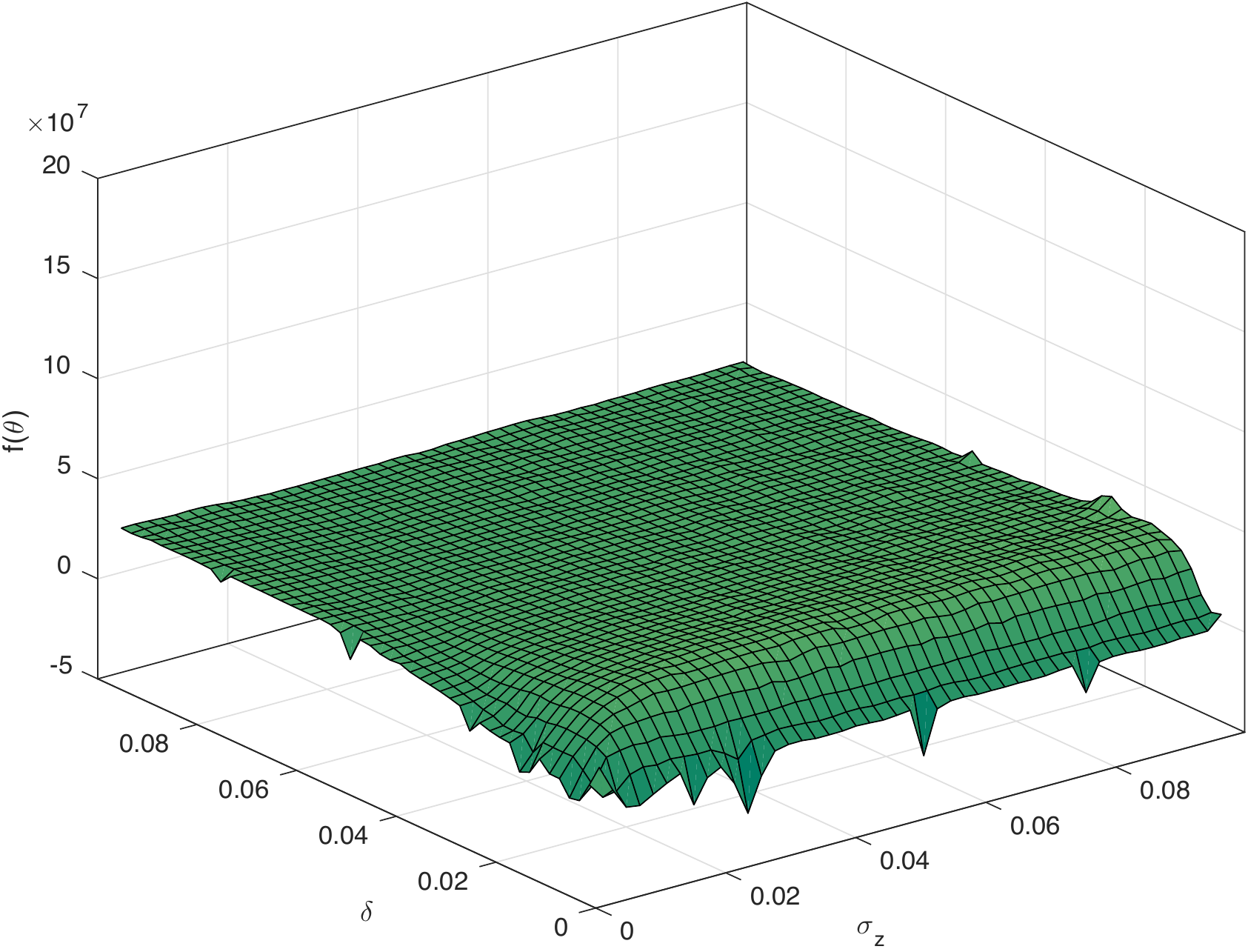}
}

\caption{Objective function surface demonstrating type 3 behavior: $\delta$ and $\sigma^z$ \label{Surface_PriceDynamics}}
\end{figure}

Finally, referring to Figure \ref{Surface_PriceDynamics}, illustrating what we will call type 3 behavior, we see that $\delta$ demonstrates a very well-behaved effect on the objective function, resulting in a very smooth surface, far more amenable to calibration. Being the parameter that best demonstrates this kind of behavior, it is not surprising that it was the only parameter that demonstrated true convergence in the calibration experiments conducted. 

It should be noted that the effect of $\sigma^z$ is far less significant than that of $\delta$, but still appears to result in well-defined effects on the objective function, observable when $\delta$ is close to $0$. In the calibration results previously presented, $\sigma^z$ typically performed much better than the other parameters, with the exception of $\delta$, having a much smaller confidence interval, but likely did not show true convergence since the effect of $\delta$ on the objective function is extremely dominant, with $\sigma^z$ only producing a noticeable effect when $\delta$ is small, but still larger than values to which $\delta$ tends to converge in the calibration experiments.

Calibration experiments involving a fixed $\delta$ may result in convincing convergence for $\sigma^z$, but such investigations are beyond the scope of this investigation.

A key realization is the fact that the effects of the other parameters on the objective function are small in comparison to the effect of $\delta$. Parameters demonstrating type 1 behavior, as shown in Figure \ref{Surface_LFHFTraders}, tend to result in objective function values of no more than $5000$ for their worst possible combinations, absolutely insignificant in comparison to $\delta$, resulting in values in excess of $2 \times 10 ^ 7$ if chosen poorly. Though $\sigma^y$ produces a significant effect on the objective function, the worst values of $\delta$ still produce objective function values more than $10$ times that of the worst values of $\sigma^y$.

This is somewhat disconcerting, as it points to one of two possible conclusions. 

The first, which we tend to lean toward, is that model itself may be flawed in some respects. To substantiate these claims, $7$ of the $10$ free parameters we considered demonstrated what we referred to as type 1 behavior, producing haphazard and predominantly small effects on the objective function values. In contrast to this, $\delta$ and to a lesser extent $\sigma^y$ and $\sigma^z$, demonstrate far more significant effects on the objective function, which seem to have resulted in $\delta$ dominating the calibration process, leading it to be the only successfully calibrated parameter. This seems to indicate that $7$ of the parameters considered demonstrate some form of parameter degeneracy, resulting in a model that is primarily driven by a single parameter, $\delta$. This is further illustrated by the fact that similar objective function values were obtained for many of the calibration experiments conducted and that these in general resulted in a reasonable fit to the empirical data. This suggests that choosing $\delta$ well may be enough to obtain reasonable calibration, at least with regard to the moments considered. We suggest possible reasons for this in Section \ref{Discussion}.

Alternatively, it may simply be a case that the method of calibration, more specifically the weight matrix, is flawed in some respects, resulting in these observed behaviors. While we do indeed acknowledge that there are flaws in method of \citet{Fabretti13} and hence our own, the successful use of the methodology to calibrate the \citet{Farmer02} model and the fact that we were able to obtain reasonable fits to empirical data and obtain convergence in the parameter $\delta$ seems to negate these concerns. We address these issues in more detail in Section \ref{Discussion}.

\subsection{Supplementary Calibration Experiments \label{Supplementary}}

In this section, we address potential criticism relating to the approach taken to calibrate the model, such that we are able to ensure that our conclusions are robust. Specifically, we address the fact that we attempt to calibrate $10$ free parameters in the preceding calibration experiments, but employ only $5$ moments in our objective function.

In this context, it is important to realize that the work of \citet{Fabretti13} and the body of work on which it is based, namely the work of \citet{Gilli03}, similarly employs fewer moments than free parameters in the conducted estimation procedures.

Nevertheless, not addressing such criticism would place significant doubt on whether the conclusions reached in this investigation are robust or simply the result of experimental design. We therefore repeat our calibration experiments, as before, but this time consider a smaller free parameter set and apply only the Nelder-Mead simplex algorithm combined with the threshold accepting heuristic.

In these experiments, we set $N_H$, $N_L$, $\delta$ and $\sigma^z$ to be the only free parameters and verify that $\delta$ once again emerges as a more easily identifiable parameter than the remaining free parameters and that the associated objective function surface once again demonstrates type $3$ behavior. In addition to this, we verify that $N_H$ and $N_L$ are once again difficult to uniquely identify and that the associated parameter surface again demonstrates type $1$ behavior, indicative of parameter degeneracies.

\begin{table}[H]
\caption{Nelder-Mead Simplex Calibration Results (Reduced Free Parameter Set)}
\centering
\begin{tabular*}{\linewidth}{@{\extracolsep{\fill}}cccc}
\hline
Parameter & 95\% Conf Int & $\frac{s}{\sqrt{n}}$ \\
\hline
$\delta$ & $[0.0016, 0.0063]$ & $0.0011$ \\
$\sigma^z$ & $[0.0238, 0.0452]$ & $0.0051$ \\
$N_L$ & $[835, 3696]$ & $683.5941$ \\
$N_H$ & $[2082, 8935]$ & $1637.0516$ \\
\hline
\end{tabular*}

\vspace{0.2cm}
\caption*{\small 95\% confidence intervals for the reduced set of free parameters, obtained from 20 independent calibration experiments involving the Nelder-Mead simplex algorithm combined with the threshold accepting heuristic \label{SimplexResultsReduced}}
\end{table}

\vspace{-0.5cm}

\begin{table}[H]
\caption{Best Parameter Set Obtained Through Calibration (Reduced Free Parameter Set) \label{BestParametersReduced}}
\centering
\begin{tabular*}{\linewidth}{@{\extracolsep{\fill}}ccc}
\hline
Calibrated Parameter & Value\\
\hline
$\delta$ & $0.000008$ \\
$\sigma^z$ & $0.0001$ \\
$N_L$ & $1005$ \\
$N_H$ & $7299$ \\
\hline
\end{tabular*}

\vspace{0.2cm}
\caption*{\small Best parameter set obtained through the implementation of the Nelder-Mead simplex algorithm combined with the threshold accepting heuristic in experiments involving the reduced free parameter set}
\end{table}

\vspace{-0.5cm}

\begin{figure}[H]
\centering
\includegraphics[width=0.9\linewidth]{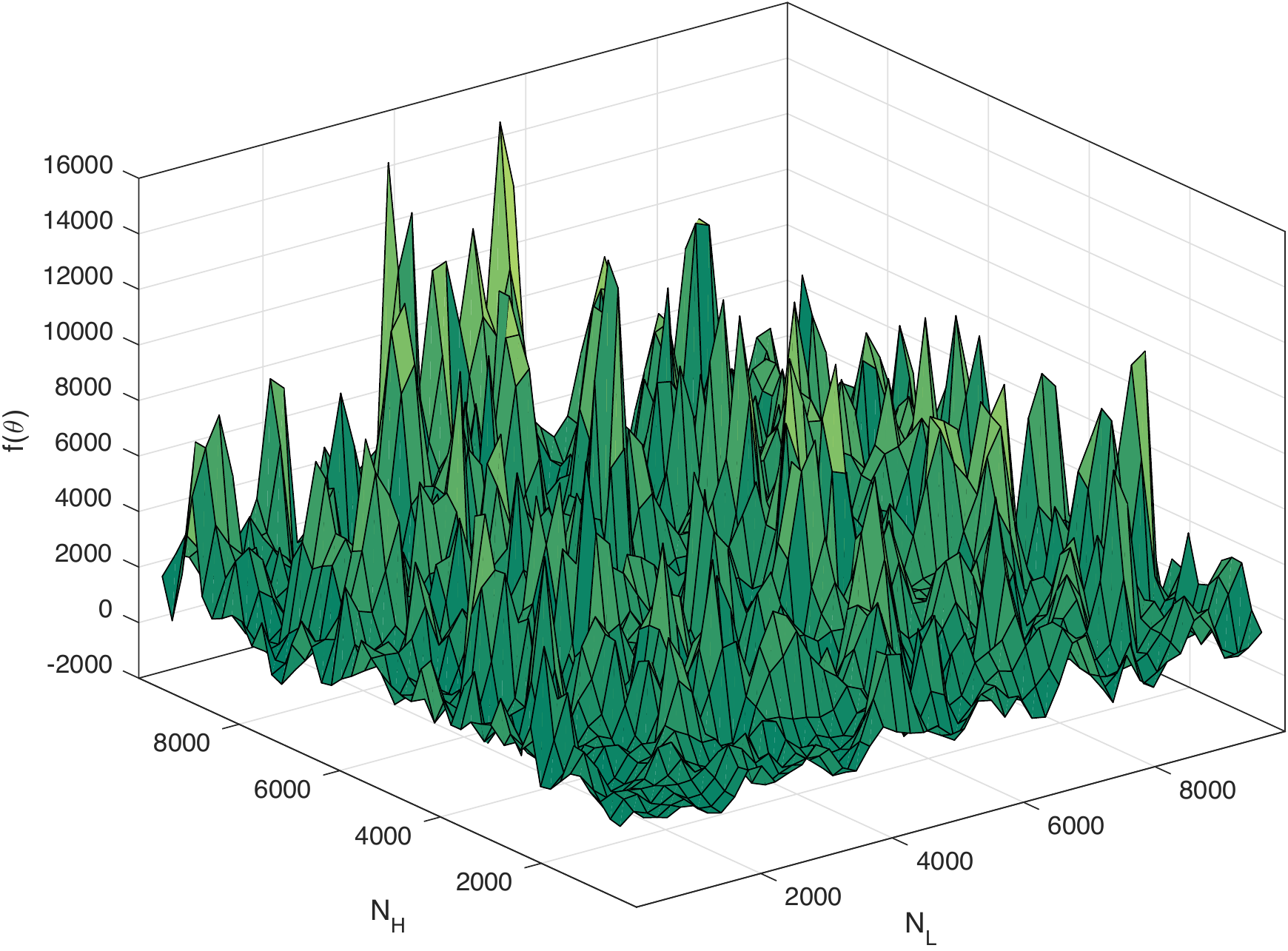}

\caption{Objective function surface illustrating type 1 behavior ($N_H$ and $N_L$), generated using the parameter values in Table \ref{BestParametersReduced} as the base values for the free parameters, with all other parameters set according to the values specified in Table \ref{JLParameters} \label{Surface_LFHFTraders_Reduced}}
\end{figure}

\begin{figure}[H]
\centering
\includegraphics[width=0.9\linewidth]{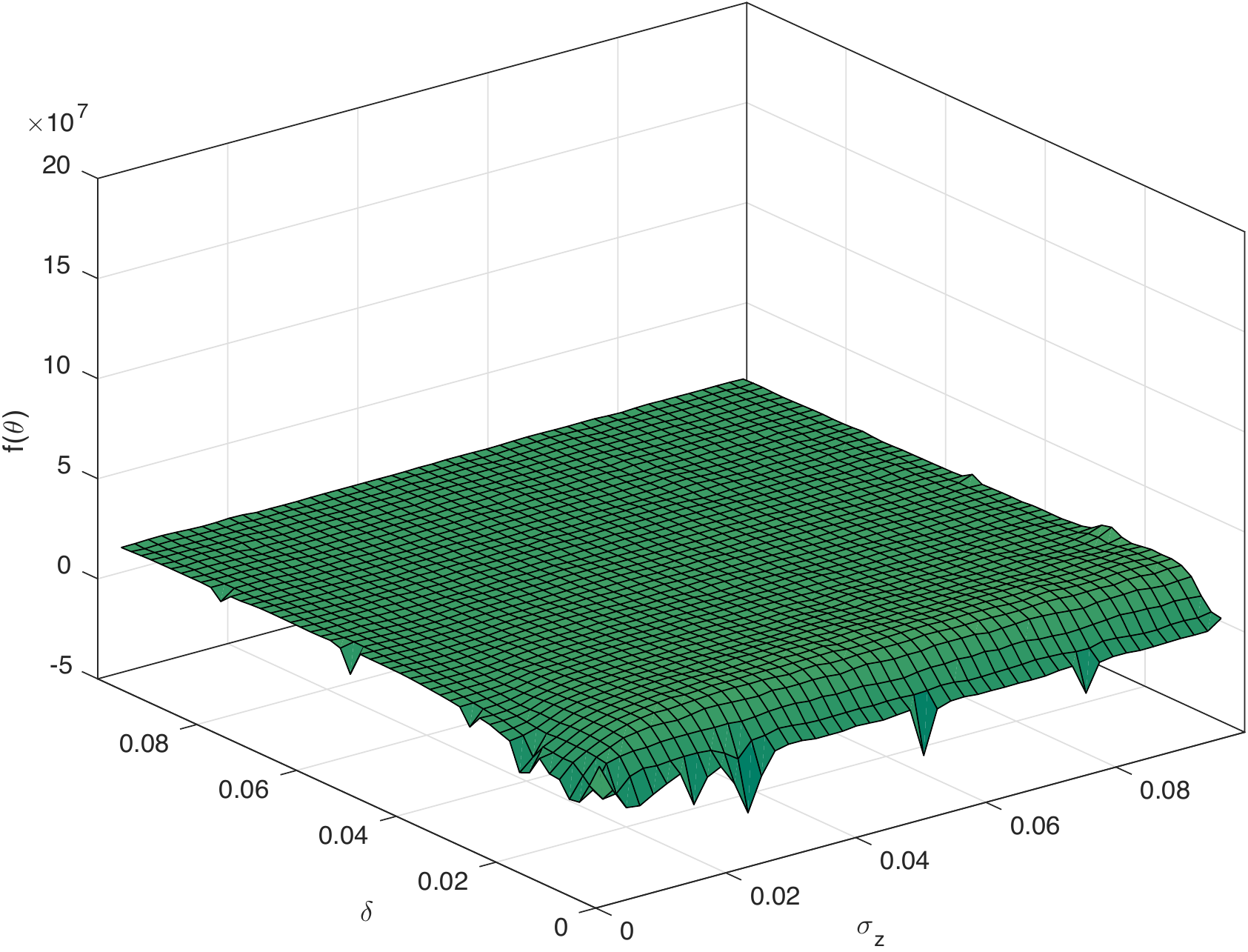}

\caption{Objective function surface illustrating type 3 behavior ($\delta$ and $\sigma^z$), generated using the parameter values in Table \ref{BestParametersReduced} as the base values for the free parameters, with all other parameters set according to the values specified in Table \ref{JLParameters} \label{Surface_PriceDynamics_Reduced}}
\end{figure}

As is clearly evident in Table \ref{SimplexResultsReduced}, we observe that $\delta$ once again emerges as showing some evidence of convergence, with a much smaller confidence interval and standard error than the remaining free parameters. Furthermore, referring to Figures \ref{Surface_LFHFTraders_Reduced} and \ref{Surface_PriceDynamics_Reduced}, we observe that $N_H$ and $N_L$ once again result in a type $1$ objective function surface, though slightly different to that observed in Figure \ref{Surface_LFHFTraders}, while $\delta$ and $\sigma^z$ once again result in a type $3$ objective function surface that is nearly identical to that presented in Figure \ref{Surface_PriceDynamics}. This is in spite of the fact that the calibration experiments were repeated with fewer free parameters than objective function moments.

It is therefore evident that our results cannot be explained as a mere consequence of the relative number of moments and parameters employed in the calibration procedures.

\section{Discussion of Results \label{Discussion}}

As was shown in Section \ref{CalResults}, $\alpha^c$, $\sigma^c$, $\alpha^f$, $\sigma^f$, $\lambda$, $N_L$ and $N_H$ demonstrate evidence of parameter degeneracies and have a haphazard effect on the objective function, while $\sigma^y$ and $\sigma^z$ present evidence of an observable objective function trend and $\delta$ demonstrates a clear relationship between its values and the objective function value. This dominance of $\delta$ on the model dynamics may be a direct result of the model's construction:

Firstly, referring to Eqn. (\ref{LFOrderPrice}), we see that LF trader order prices for a particular session are set according to a single iteration of a geometric random walk referencing the market price of the previous session. Not surprisingly, $\delta$ is an important parameter in determining the nature of the aforementioned geometric random walk and hence $\delta$ plays an important role in the setting of LF trader order prices in the model. Furthermore, HF trader orders expire at the end of each session ($\gamma^H$ = 1) and HF trader order prices are set to be slight perturbations of existing LF trader orders in the order book, hence we see that $\delta$ would have a significant impact on HF trader prices in each session as well.

Secondly, referring to Eqn. (\ref{Fundamentalist}), we see that both $\delta$ and $\sigma^y$, another parameter shown to demonstrate structure in the calibration experiments, drive the equation which determines the fundamental value. Because order sizes under both the chartist and fundamentalist strategies are predominantly determined by the market price and fundamental value time series, both of which are driven by or heavily influenced by geometric random walks including $\delta$ as a parameter, we see that $\delta$ has a significant effect on both the model's order size and price dynamics. This is in contrast to the remaining parameters, which typically only influence one aspect of the model. This leads to a fairly significant coupling of dynamics in the model, making $\delta$ far more important than the remaining model parameters and possibly leading it to dominate the calibration experiments as it has. \citet{Chakraborti11} state that it is often difficult to identify the role played by and dependencies between the various parameters present in ABMs, despite their ability to reproduce well-known stylized facts. Our investigation has demonstrated a similar finding.

We see that $\sigma^y$, $\sigma^z$ and $\delta$, the only parameters demonstrating any structured effect on the objective function in the calibration experiments, appear in two geometric random walks included in the model's equations. It may be that these random walk dynamics dominate the model, with the parameters not considered in these random walks degenerating to noise. It was previously shown in Section \ref{CalResults}.\ref{ParamAnalysis} that the effects of $\sigma^y$, $\sigma^z$ and $\delta$ on the objective function are far more significant than that of the remaining parameters, again pointing to a similar conclusion. This suggests that the inclusion of a random walk may be a far stronger and dominant model assumption than one may perceive it to be at first glance. 

These observed degeneracies may also find their origin in the more realistic matching processes used to determine market prices in intraday ABMs of double auction markets, in contrast to the approximating equations used in daily models, which have been more successfully calibrated by \citet{Fabretti13} and \citet{Gilli03}. Perhaps the indirect calculation of prices through order matching does not lead to the same well-defined dynamics as an approximating equation that considers the effects of all parameters with simple assumptions regarding their relationship? It is also worth considering why only parameters dealing with order price dynamics, $\delta$ and $\sigma^z$, resulted in a smooth objective function surface.

The exact nature and cause of these degeneracies requires further investigation and would likely require the consideration of other models also involving realistic order matching processes, but with a different approach to that of \citet{JacobLeal15}, since it is possible that degeneracies may emerge due to the current LF trader order price mechanism. 

As an example, one might consider the \citet{Preis06} model, also involving realistic matching processes in a continuous double auction market, but using a market microstructure-based order placement mechanism as opposed to a random walk determining order prices. In general, this type of order placement mechanism is more desirable, since the use of a random walk by \citet{JacobLeal15} is unintuitive and results in orders being placed without any consideration of the dynamics of the order book. Considering the importance of price impact in contemporary literature \citep{Lehalle13, Cartea15}, order placement behaviors more directly rooted in the dynamics of the order book are worth investigating.

It is evident through the calibration experiments conducted that the model or perhaps even this particular class of continuous double auction models present a number of obstacles that render them difficult to calibrate definitively. This, in essence, presents a number of challenging questions. How can we be sure that the parameter $N_H$ truly influences the nature of HF trading in a simulated market rather than introducing noise into a random walk model? Furthermore, how can we trust the conclusions drawn by \citet{JacobLeal15} regarding the parameter $N_H$ and the prevalence of flash crashes, when $N_H$ seems to be a degenerate parameter? It should also be noted that Figure \ref{Surface_LFHFTraders}, demonstrating the presence of many vastly different combinations of $N_L$ and $N_H$ resulting in similar objective function values, suggests that answering the question raised at the start of this investigation, namely the extent to which HF trading strategies have been adopted in South Africa, is essentially impossible, as the values obtained for $N_L$ and $N_H$ demonstrate almost no structure in general. 

Being unable to obtain unique calibrations for a particular model rigorously is in fact a more serious issue than it is presented as in much of the literature. Without a unique set of parameters allowing us to reproduce the properties of financial time series drawn from a particular market, how can we argue that introducing and observing changes in the model truly reflects the same changes in the market?

It is important, however, to remember that we have attempted to calibrate the model to a single week of data for a particular stock traded on a particular market. Therefore, the above are not definitive conclusions, but rather discussions related to potential calibration challenges identified by this particular investigation. In future, it would certainly be worth applying similar methodologies to other intraday datasets to verify whether this behavior is common or isolated. It may very well be that other data sets yield more well-defined behaviors with respect to important parameters, such as $N_H$.

Of course, it is entirely possible, but in our opinion unlikely, that the aforementioned concerns result from the calibration methodology and not the model itself. Firstly, it must be noted that the weight matrix, shown in Appendix \ref{WeightMatrix}, was obtained by inverting a matrix that is non-singular, but only just. \citet{Fabretti13} also performs the aforementioned inversion of a near-singular matrix, though we inverted a less ill-conditioned matrix. This could of course be used to explain the haphazard behaviors demonstrated by some of the parameters, but if and only if this was a general trend. As we have shown, this same weight matrix results in fairly well-defined behaviors with regard to the parameter $\delta$, making this argument harder to justify. It should be noted that the calibration experiments did not fail as such. While most of the parameters could not be uniquely identified, $\delta$ showed evidence of convergence and the obtained fits of the model to the empirical data, shown in Section \ref{CalResults}.\ref{DataComparison}, were in fact relatively good, more or less in line with that of \citet{Fabretti13}, who was considered to have been successful. It is also worth noting that we have not deviated from the methodology of \citet{Fabretti13}, regarded as a valid calibration approach.

In addition to the above, we have also addressed the criticism that the number of free parameters considered in the calibration experiments exceeds the number of moments used to construct the objective function in Section \ref{CalResults}.\ref{Supplementary}. If, in the continuous time limit and under the assumptions of ergodicity and stationarity, the number of moments in the objective function is at least the same as the number of parameters in model to be calibrated, then there can be perfect model parameter identification and the parameter values found are the solution that minimizes the objective function \citep{Canova09}. This is not in general possible, in part due to the large number of parameters characteristic of the class of models considered.

Under-identified moment estimators must be approached with care \citep{Canova09}, however, the problem of estimation of an ABM can reside with the model itself rather than the estimation method when a subset of the parameters can be well-identified -- a subset with a clear and specific physical interpretation. This may imply that calibration (which is weaker than estimation because of the auxiliary assumptions made) can be used to determine whether a model is appropriate or not. The model we have investigated demonstrates degeneracies even when considering fewer free parameters than moments in separate experiments (where the same subset of parameters remained identifiable), which we consider sufficient to negate concerns related to over- and under-identification \citep{Bollen89, Rigdon95, Canova09} and sufficient to question the model specification, even though the model is able to recover various stylized facts.

Regardless of one's stance on whether the behavior observed in this investigation is isolated, or indicative of general trends in rules-based ABMs involving realistic order-matching processes at an intraday time scale, we have demonstrated a concern. While Section \ref{StylizedFacts} demonstrates the ability of our implementation of the model to replicate a number of established log return time series stylized facts, this did not account for the fact that the model demonstrated evidence of parameter degeneracies and appeared to be driven predominantly by a single parameter, with much of the remaining processes tending towards noise. This is testament to a problematic oversight that permeates the breadth of much of the contemporary literature on the subject: the sufficiency of stylized facts as a form of validation. The perceived sufficiency of stylized facts as a form of validation is so pervasive that \citet{Panayi13} suggest it is more or less standard practice in financial agent-based modeling. While a model must replicate stylized facts to be validated, the replication of stylized facts can only be taken as a first step and successful calibration to transaction data must be achieved to truly validate a model. 

Despite a number of shortcomings, ABMs still have many important contributions to offer. In particular, the effect of HF trader activity on the stability and dynamics of financial markets is still being debated \citep{Easley12} and ABMs may hold the answers to many unanswered questions. Other modeling approaches, such as those employed by \citet{Gerig12}, have been used to show the price synchronizing features of HF trading, as well as features leading to improvements in market efficiency during normal periods and negative consequences during times of stress. ABMs are well placed to contribute similarly to this debate.

In future work, it may very well arise that rules-based, intraday ABMs involving realistic matching processes may all demonstrate similar flaws when subjected to rigorous calibration. For this reason, it would be necessary to mention possible alternatives that may represent a step up from overly-simplified daily closed-form models. 

One such alternative is to revisit the mathematics of trader rules and consider the replacement of the disjointed order price, order size and trader activation rules with a single structure, namely a Hawkes process, to generate trade events. In recent years, much progress has been made in using Hawkes processes to generate realistic occurrences of order book events \citep{Bacry15, Martins16}. Hawkes processes may also allow the model to function more appropriately in an event-based paradigm \citep{Hendricks16b}, possibly negating a need to down-sample event-based tick data into a chronological series, which may have resulted in the problems previously observed. An example of the incorporation of Hawkes process into agent-based order book simulation is presented by \citet{Toke11}. 

Alternatively, it may be advantageous to dispense with rules entirely, focusing instead on large scale simulations involving many purposeful, reinforcement-learning agents with no initially assumed rules \citep{Hendricks14, Hendricks16a, Wilcox14} and focus on competing, adaptive agents in a game theoretic framework \citep{Galla13, Lachapelle15}. This may require a more effective adoption of a large-scale multi-agent system (MAS) approach for the modeling of ABM systems for market simulation applications\footnote{The authors benefited from conversations with D. Pugh and J.D. Farmer in relation to the large-scale simulation of purposeful agents in the financial market setting.}, which would require appropriately reactive software platforms, such as Akka \citep{Wampler15}. This in turn would create new challenges for model validation. 

\section{Conclusion}

We have provided an example of a financial ABM that is capable of recovering a number of well-known stylized facts of return time series, while also demonstrating parameter degeneracies and calibration difficulties. This casts doubt on the meaningfulness of the stylized fact-centric ABM model validation prevalent in much of the contemporary literature as it relates to intraday ABMs approximating continuous double auction market phenomenology and dynamics. Furthermore, the observed parameter degeneracies make it extremely challenging to accept any insight gained through varying parameter values and observing changes in model behavior.

This work does not serve to undermine the extensive and predominantly successful work that has been done in the domain of financial agent-based modeling, particularly with regards to closed-form daily ABMs, but rather to caution model developers against the blind acceptance of stylized fact-centric validation, which can, as it has here, fail to identify fundamental model problems.

We suggest that future model validation processes should instead focus on demonstrating the ability of an ABM to reproduce the statistical properties of empirically measured transaction data through calibration, while still including stylized-fact verification as a separate component of model validation.

Accidental agreements between models and stylized facts need to be guarded against. The pragmatic calibration approach of \citet{Fabretti13} provides an additional probe of model structure in the context of continuous double-auction markets, one that may further help balance the tenants of verisimilitude and model corroboration with the conceptual benefits of ABMs.

\section{Acknowledgements}
 
T.J. Gebbie acknowledges the financial support of the National Research Foundation (NRF) of South Africa (grant number 87830). D.F. Platt acknowledges the financial support of the University of the Witwatersrand, Johannesburg and NRF of South Africa. The conclusions herein are due to the authors and the NRF accepts no liability in this regard. 

T.J. Gebbie would like to thank the INET-Oxford Complexity Economics research group for hospitality and stimulating interactions on the application of MAS and ABM models in finance. T.J. Gebbie and D.F. Platt would like to thank S. Jacob Leal for helpful conversations relating to agent-based modeling.

\end{multicols}

\section*{Appendices}

\appendix

\begin{multicols}{2}

\subsection{Nelder-Mead Simplex Algorithm \label{SimplexDescription}}

The \citet{Nelder65} simplex algorithm is based upon the fundamental idea that the current solution consists of the points of a simplex with $n + 1$ vertices in the parameter space, $\Theta$, where $n$ is the number of parameters being calibrated in our context. During each iteration, we calculate the value of $f$, the objective function, at the vertices of the simplex and consider the vertex with the worst objective function value. We then consider the reflection, expansion, in-contraction and out-contraction points of this vertex. If this leads to a point where the value of the objective function is improved, we update the vertex with the worst objective function value, but if none of these points leads to an improvement, the simplex shrinks. We consider the version of the algorithm presented by \citet{Gilli03} and perform $5$ Monte Carlo replications when applying the method.

\noindent\rule{7.5cm}{0.4pt}

\vspace{-0.15cm}
\noindent \textbf{Algorithm 1}: Nelder-Mead Simplex Algorithm

\vspace{-0.36cm}
\noindent\rule{7.5cm}{0.4pt}

\begin{algorithmic}[1] \label{Simplex}
\WHILE{stopping criteria not met}
\STATE{Rename vertices such that $f(\theta^{(1)})\leq...\leq f(\theta^{(n + 1)})$}
\IF{$f(\theta^R) < f(\theta^{(1)})$}
\IF{$f(\theta^E) < f(\theta^{R})$} 
\STATE{$\theta^* = \theta^E$} 
\ELSE 
\STATE{$\theta^* = \theta^R$} 
\ENDIF
\ELSE
\IF{$f(\theta^R) < f(\theta^{(n)})$}
\STATE{$\theta^* = \theta^R$} 
\ELSE
\IF{$f(\theta^R) < f(\theta^{(n + 1)})$}
\IF{$f(\theta^O) < f(\theta^{(n + 1)})$}
\STATE{$\theta^* = \theta^O$}
\ELSE
\STATE{shrink}
\ENDIF
\ELSE
\IF{$f(\theta^I) < f(\theta^{(n + 1)})$}
\STATE{$\theta^* = \theta^I$}
\ELSE
\STATE{shrink}
\ENDIF 
\ENDIF
\ENDIF
\ENDIF
\IF{not shrink}
\STATE{$\theta^{(n + 1)} = \theta^*$}
\ENDIF 
\ENDWHILE
\end{algorithmic}

\vspace{-0.2cm}
\noindent\rule{7.5cm}{0.4pt}

\begin{table}[H]
\caption{Nelder-Mead Simplex Vertex Transformations}
\begin{tabular}{cc}
\hline
Operation & Equation \\ \hline
Mean & $\bar{\theta} = \frac{1}{n} \sum_{i=1}^{n}\theta^{(i)}$ \\
Reflection & $\theta^R = (1 + \rho) \bar{\theta} - \rho \theta^{(n + 1)}$ \\
Expansion & $\theta^E = (1 + \rho \xi) \bar{\theta} - \rho \xi \theta^{(n + 1)}$ \\
Out-contraction & $\theta^O = (1 + \psi \rho) \bar{\theta} - \psi \rho \theta^{(n + 1)}$ \\
In-contraction & $\theta^I = (1 - \psi \rho) \bar{\theta} + \psi \rho \theta^{(n + 1)}$ \\ 
Shrink simplex & $\theta^{(i)} = \theta^{(1)}-\sigma(\theta^{(i)} - \theta^{(1)})$, \\
toward $\theta^{(1)}$ & $i = 2,...,n + 1$ \\ \hline
\end{tabular}
\vspace{0.2cm}
\caption*{Computation of transformed points as required by the Nelder-Mead simplex algorithm. We set $\rho = 1$, $\xi = 2$, $\psi = \frac{1}{2}$ and $\sigma = \frac{1}{2}$, as is the standard for the algorithm \citep{Gao10}}
\end{table}

\vspace{-0.5cm}

While the Nelder-Mead simplex algorithm is relatively robust in the case of a smooth objective function, it tends to converge to local minima when the objective function being considered is not smooth. \citet{Gilli03} highlight this problem in the context of agent-based modeling, as typical objective functions associated with agent-based modeling problems are usually not smooth. For this reason, the Nelder-Mead simplex algorithm is combined with the threshold accepting heuristic to aid convergence to a global minimum.

In threshold accepting, we randomly perturb a set of parameters and accept new parameter sets which are not worse than our current best set, as measured by the objective function, by more than a certain threshold, $\tau$. We typically have a set of $n_R$ thresholds, $\tau_r, r = 1,...,n_R$. When combining this heuristic with the Nelder-Mead simplex algorithm, we begin by shifting the current simplex in a random direction, with the magnitude of this shift also being random. We then accept this new simplex if the vertex with the best objective function value in the new simplex has an objective function value less than the best in the previous simplex, after the threshold has been added.

We implement the required random shift as follows: 
\begin{enumerate}
\item We begin by \textit{randomly selecting} one of the free parameters considered in the optimization problem.
\item We then \textit{multiply the mean value} of this parameter across the simplex vertices by a random number drawn from $\mathcal{U}(-0.5, 0.5)$ and add this quantity to the current value of the selected parameter at each vertex.
\end{enumerate}

We perform $n_R = 3$ rounds in each application of threshold accepting, each consisting of $n_s = 5$ steps, with $\tau_1 = \frac{1}{5}$, $\tau_2 = \frac{1}{10}$ and $\tau_3 = 0$. We perform $3$ Monte Carlo replications in round $1$, $4$ in round $2$ and $5$ in Round $3$.

A single step in the combined algorithm \citep{Gilli03}, involving both the Nelder-Mead simplex algorithm and threshold accepting heuristic, is described as follows:

\noindent\rule{7.5cm}{0.4pt}

\vspace{-0.1cm}
\noindent \textbf{Algorithm 2}: Combined Algorithm (NM + TA)

\vspace{-0.3cm}
\noindent\rule{7.5cm}{0.4pt}

\begin{algorithmic}[1]
\STATE{Generate $u$ from $\mathcal{U}(0, 1)$}
\IF{$u < \xi$}
\FOR{$r = 1$ to $n_R$}
\FOR{$s = 1$ to $n_s$}
\STATE{Generate $\theta_{new} \in \mathcal{N}_{\theta_{old}}$ using a random shift}
\IF{$f(\theta_{new}) < f(\theta_{old}) + \tau_r$}
\STATE{Accept $\theta_{new}$}
\ENDIF
\ENDFOR
\ENDFOR
\ELSE
\STATE{Generate $\theta_{new} \in \mathcal{N}_{\theta_{old}}$ with a simplex search and check for acceptance}
\ENDIF
\IF{$\theta_{new} $ is accepted}
\STATE{$\theta_{old} = \theta_{new}$}
\ENDIF
\end{algorithmic}

\vspace{-0.2cm}
\noindent\rule{7.5cm}{0.4pt}

In the above, we set $\xi = 0.15$.

\subsection{Genetic Algorithm \label{GADescription}}

We refer to the class of algorithms proposed by \citet{Holland75}. In essence, genetic algorithms are parameter selection and perturbation strategies, inspired by evolutionary processes in nature, used to solve optimization problems. Like organisms in the natural world, superior solutions survive and pass on characteristics, whereas inferior solutions tend to become extinct.

The basic principles of genetic algorithms, as summarized by \citet{Whitley94}, are given as follows:
\begin{enumerate}
\item We begin with an \textit{initial population}, usually randomly generated, consisting of possible solutions to the optimization problem, which are called chromosomes in the context of the literature.
\item Chromosomes that represent better solutions to the target problem (as measured by a fitness function), are given \textit{better opportunities} to reproduce (crossover) as compared to inferior chromosomes.
\item \textit{Reproduction} (crossover) ultimately results in a new population that hopefully contains improved chromosomes, in other words, those that result in lower (higher) objective function values in the context of a minimization (maximization) problem.
\end{enumerate}
Genetic algorithms have become increasingly prevalent in contemporary literature, a result of the fact that they are both intuitive and relatively robust. They do not require gradient information and are thus well suited to functions with many local minima (or maxima) and non-differentiable functions. For this reason, genetic algorithms are typically widely applied, unless an applicable gradient or problem-specific method exists \citep{Whitley94}. Considering that the constructed objective function in our framework does indeed have many local minima and considering the lack of any specialized algorithms for ABM calibration, a genetic algorithm seems to be a viable candidate.

We consider the Simple Genetic Algorithm (SGA), described by \citet{Goldberg89} as follows:

\vspace{0.2cm}

\noindent\rule{7.5cm}{0.4pt}

\vspace{-0.1cm}
\noindent \textbf{Algorithm 3}: Simple Genetic Algorithm

\vspace{-0.3cm}
\noindent\rule{7.5cm}{0.4pt}

\begin{algorithmic}[1]
\STATE{Randomly initialize the population \footnote{We set the population size to be $100$ and consider $5$ Monte Carlo replications per individual.}}
\STATE{Determine the fitness values of the current population}
\WHILE{best individual in the population does not meet the required tolerance}
\STATE{\textbf{Select} parents at random to create an intermediate population}
\STATE{Perform a \textbf{crossover} operation on the parents to generate offspring}
\STATE{Apply the \textbf{mutation} operation to the population of offspring}
\STATE{Calculate the fitness values for the new population}
\ENDWHILE
\end{algorithmic}

\vspace{-0.2cm}
\noindent\rule{7.5cm}{0.4pt}

\vspace{0.2cm}

The genetic operators highlighted in the above algorithm are discussed in detail as follows:
\begin{enumerate}
\item During a \textit{selection} operation, chromosomes are randomly selected with replacement and probability proportional to their current fitness values to create an intermediate population.
\item During a \textit{crossover} operation, chromosomes are first randomly paired for recombination. Since the selection process sufficiently shuffles the population, the pairs are simply set to be the first and second chromosomes in the intermediate population, the third and fourth chromosomes in the intermediate population and so on. Thereafter, the strings recombine, with probability $p_c$, to form new chromosomes that constitute the new population. The specific recombination process is dependent on the implementation.
\item During the \textit{mutation} operation, a slight perturbation is applied to the population, with a very small probability, $p_m$. This typically involves the slight shifting of a single parameter value associated with a single chromosome in the population.
\end{enumerate}

It should be noted that fitness and objective functions need not be the same. Fitness functions typically measure the performance of a chromosome relative to the population whereas objective functions tend to measure the performance of the chromosome in the context of the optimization problem itself. In our case, however, we set the fitness function to be identical to the objective function defined for the calibration framework.

We make use of MATLAB's \textit{ga} function, an implementation of the above algorithm, packaged as part of the \textit{Global Optimization Toolbox}. The problem encoding as well as the crossover and mutation operators in our context are defined as per the defaults for this function. Parallel objective function evaluations are performed using a master-slave paradigm, similar to that implemented by \citet{Hendricks16c}.

\end{multicols}

\subsection{Weight Matrix Required by the Method of Simulated Moments \label{WeightMatrix}}

The weight matrix below corresponds to the data set representing transaction data for Anglo American PLC over the period beginning at 9:10 on 1 November 2013 and ending at 16:50 on 5 November 2013. The diagonal elements correspond to the mean, standard deviation, kurtosis, KS test statistic and Hurst exponent respectively.

\begin{equation*}
\mathbf{W} = 
\begin{pmatrix}
5.0346 \times 10 ^ 4 & -1.2885 \times 10 ^ 4 & -736.6343 & 3.0220 \times 10 ^ 3 & 391.2534 \\
-1.2885 \times 10 ^ 4 & 7.8957 \times 10 ^ 5 & 2.5435 \times 10 ^ 3 & -341.4378 & -6.1999 \times 10 ^ 3 \\
-736.6343 & 2.5435 \times 10 ^ 3 & 28.7473 & -88.4746 & 17.2640 \\
3.0220 \times 10 ^ 3 & -341.4378 & -88.4746 & 723.4611 & 56.7301 \\
391.2534 & -6.1999 \times 10 ^ 3 & 17.2640 & 56.7301 & 2.3549 \times 10 ^ 3 \\
\end{pmatrix}
\end{equation*}

It should be noted that the above weight matrix is similar to that obtained by \citet{Fabretti13}. Our weight matrix is obtained by inverting $Cov[\mathbf{m}^e]$, which had a condition number of $1.2772 \times 10 ^ 5$. While this may seem problematic, it must be noted that \citet{Fabretti13} inverts $Cov[\mathbf{m}^e]$ with a condition number of $2.6964 \times 10 ^ 5$ to obtain the weight matrix, indicating that our weight matrix is at least as stable.


\begin{thebibliography}{99}

\bibitem[Alfarano et al.(2005)]{Alfarano05}
Alfarano S, Lux T, Wagner F (2005) Estimation of agent-based models: the case of an asymmetric herding model. Comput Econ 26(1):19-49

\bibitem[Alfarano et al.(2006)]{Alfarano06}
Alfarano S, Lux T, Wagner F (2006) Estimation of a simple agent-based model of financial markets: an application to Australian stock and foreign exchange data. Physica A 370(1):38-42

\bibitem[Alfarano et al.(2007)]{Alfarano07}
Alfarano S, Lux T, Wagner F (2007) Empirical validation of stochastic models of interacting agents. Eur Phys J B 55(2):183-187

\bibitem[Amilon(2008)]{Amilon08}
Amilon H (2008) Estimation of an adaptive stock market model with heterogeneous agents. J Empir Financ 15(2):342-362

\bibitem[Bacry et al.(2015)]{Bacry15}
Bacry E, Mastromatteo I, Muzy JF (2015) Hawkes processes in finance. arXiv:1502.04592

\bibitem[Barde(2016)]{Barde16}
Barde S (2016) Direct calibration and comparison of agent-based herding models of financial markets. J Econ Dyn Control 73:329-353

\bibitem[Bollen(1989)]{Bollen89}
Bollen KA (1989) Structural Equations with Latent Variables. John Wiley \& Sons, Hoboken

\bibitem[Bouchaud et al.(2002)]{Bouchaud02}
Bouchaud JP, Mezard M, Potters M (2002) Statistical properties of stock order books: empirical results and models. Quant Financ 2(4):251-256

\bibitem[Canova and Sala(2009)]{Canova09}
Canova F, Sala L (2009) Back to square one: Identification issues in DSGE models. J Monetary Econ 56:431-449

\bibitem[Cartea et al.(2015)]{Cartea15}
Cartea A, Jaimungal S, Penalva J (2015) Algorithmic and High-Frequency Trading. Cambridge University Press, Cambridge

\bibitem[Chakraborti et al.(2011)]{Chakraborti11}
Chakraborti A, Toke IM, Patriarca M, Abergel F (2011) Econophysics review: II. agent-based models. Quant Financ 11(7):1013-1041

\bibitem[Chiarella et al.(2009)]{Chiarella09}
Chiarella C, Iori G, Perello J (2009) The impact of heterogeneous trading rules on the limit order book and order flows. J Econ Dyn Control 33:525-537

\bibitem[Cont(2001)]{Cont01}
Cont R (2001) Empirical properties of asset returns: stylized facts and statistical issues. Quant Financ 1:223-236

\bibitem[Cont et al.(1997)]{Cont97}
Cont R, Potters M, Bouchaud JP (1997) Scaling in stock market data: stable laws and beyond. arXiv:9705087

\bibitem[Cooley(1997)]{Cooley97}
Cooley T (1997) Calibrated models. Oxf Rev Econ Policy 13(3):55-69

\bibitem[Di Matteo(2007)]{DiMatteo07}
Di Matteo T (2007) Multi-scaling in finance. Quant Financ 7:21-36

\bibitem[Easley et al.(2012)]{Easley12}
Easley D, Lopez de Prado M, O'Hara M (2012) The Volume Clock: Insights into the High-Frequency Paradigm. J Portfolio Manage 39:19-29

\bibitem[Fabretti(2013)]{Fabretti13}
Fabretti A (2013) On the problem of calibrating an agent-based model for financial markets. J Econ Interact Coord 8:277-293

\bibitem[Farmer and Joshi(2002)]{Farmer02}
Farmer JD, Joshi S (2002) The price dynamics of common trading strategies. J Econ Behav Organ 49:149-171

\bibitem[Galla and Farmer(2013)]{Galla13}
Galla T, Farmer JD (2013) Complex dynamics in learning complicated games. Proc Natl Acad Sci USA 110:1232-1236

\bibitem[Gao and Han(2010)]{Gao10}
Gao F, Han L (2010) Implementing the Nelder-Mead simplex algorithm with adaptive parameters. Comput Optim Appl 51:259-277

\bibitem[Gerig(2012)]{Gerig12}
Gerig A (2012) High-Frequency Trading Synchronizes Prices in Financial Markets. arXiv:1211.1919

\bibitem[Gilli and Winker(2003)]{Gilli03}
Gilli M, Winker P (2003) A global optimization heuristic for estimating agent based models. Comput Stat Data Anal 42:299-312

\bibitem[Goldberg(1989)]{Goldberg89}
Goldberg DE (1989) Genetic Algorithms in Search, Optimization and Machine Learning. Addison-Wesley Longman, Boston

\bibitem[Grazzini (2012)]{Grazzini12}
Grazzini J (2012) Analysis of the Emergent Properties: Stationarity and Ergodicity. J Artif Soc Soc Simulat 15(2):7

\bibitem[Grazzini and Richiardi (2015)]{Grazzini15a}
Grazzini J, Richiardi M (2015) Estimation of ergodic agent-based models by simulated minimum distance. J Econ Dyn Control 51:148-165

\bibitem[Grazzini et al.(2015)]{Grazzini15b}
Grazzini J, Richiardi M, Tsionas M (2015) Bayesian estimation of agent-based models. Nuffield College, Economic Working Papers Series, 2015-W12

\bibitem[Guerini and Moneta(2016)]{Guerini16}
Guerini M, Moneta A (2016) A Method for Agent-Based Models Validation. Working Paper Series, Institute for New Economic Thinking, 42 

\bibitem[Hamill and Gilbert(2016)]{Hamill16}
Hamill L, Gilbert N (2016) Agent-Based Modelling in Economics. John Wiley \& Sons, Chichester

\bibitem[Harvey et al.(2017)]{Harvey17}
Harvey M, Hendricks D, Gebbie T, Wilcox D (2017) Deviations in expected price impact for small transaction volumes under fee restructuring. Physica A 471:416-426

\bibitem[Hendricks(2016)]{Hendricks16a}
Hendricks D (2016) Using real-time cluster configurations of streaming asynchronous features as online state descriptors in financial markets. arXiv:1603.06805

\bibitem[Hendricks et al.(2016a)]{Hendricks16b}
Hendricks D, Gebbie T, Wilcox D (2016a) Detecting intraday financial market states using temporal clustering. Quant Financ 16(11):1657-1678

\bibitem[Hendricks et al.(2016b)]{Hendricks16c}
Hendricks D, Gebbie T, Wilcox D (2016b) High-speed detection of emergent market clustering via an unsupervised parallel genetic algorithm. S Afr J Sci 122(1/2)

\bibitem[Hendricks and Wilcox(2014)]{Hendricks14}
Hendricks D, Wilcox D (2014) A reinforcement learning extension to the Almgren-Chriss model for optimal trade execution. Computational Intelligence for Financial Engineering \& Economics, 2014 IEEE Conference, pp 457-464

\bibitem[Heylighen(2008)]{Heylighen08}
Heylighen F (2008) Complexity and Self-organization. In: Bates MJ, Maack MN (eds) Encyclopedia of Library and Information Sciences. CRC Press, Boca Raton

\bibitem[Holland(1975)]{Holland75}
Holland JH (1975) Adaptation in Natural and Artificial Systems. University of Michigan Press, Ann Arbor

\bibitem[Jacob Leal et al.(2015)]{JacobLeal15}
Jacob Leal S, Napoletano M, Roventini A, Fagiolo G (2015) Rock Around the Clock: An Agent-Based Model of Low- and High-frequency Trading. J Evol Econ 25:1-25

\bibitem[Kirman(1991)]{Kirman91}
Kirman A (1991) Epidemics of opinion and speculative bubbles in financial markets. In: Taylor M (ed) Money and Financial Markets. Blackwell, Oxford, pp 354-368

\bibitem[Kukacka and Barunik (2016)]{Kukacka16}
Kukacka J, Barunik J (2016) Estimation of Financial Agent-Based Models with Simulated Maximum Likelihood. SSRN:2783663

\bibitem[Kunsch(1989)]{Kunsch89}
Kunsch HR (1989) The jackknife and the bootstrap for general stationary observations. Ann Stat 17:1217-1241

\bibitem[Lachapelle et al.(2015)]{Lachapelle15}
Lachapelle A, Lasry JM, Lehalle CA, Lions PL (2015) Efficiency of the Price Formation Process in Presence of High Frequency Participants: a Mean Field Game Analysis. arXiv:1305.6323

\bibitem[Lamperti(2015)]{Lamperti15}
Lamperti F (2015) An Information Theoretic Criterion for Empirical Validation of Time Series Models. LEM Papers Series, Laboratory of Economics and Management, Sant'Anna School of Advanced Studies, 2-2015

\bibitem[Lamperti(2016)]{Lamperti16}
Lamperti F (2016) Empirical Validation of Simulated Models through the GSL-div: an Illustrative Application. LEM Papers Series, Laboratory of Economics and Management, Sant'Anna School of Advanced Studies, 18-2016

\bibitem[LeBaron(2005)]{LeBaron05}
LeBaron B (2005) Agent-based Computational Finance. In: Judd KL, Tesfatsion L (eds) The Handbook of Computational Economics, Vol. 2. Elsevier, Amsterdam, pp 1187-1233

\bibitem[Lehalle(2013)]{Lehalle13}
Lehalle CA (2013) Market Microstructure Knowledge Needed for Controlling an Intra-Day Trading Process. arXiv:1302.4592

\bibitem[Macal and North(2010)]{Macal10}
Macal CM, North MJ (2010) Tutorial on agent-based modelling and simulation. J Simulat 4:151-162

\bibitem[Mandes(2015)]{Mandes15}
Mandes A (2015) Impact of inventory-based electronic liquidity providers with a high frequency event and agent-based modeling framework. Joint Discussion Paper Series in Economics, Department of Business Administration and Economics, University of Marburg, 15-2015

\bibitem[Martins and Hendricks(2016)]{Martins16}
Martins R, Hendricks D (2016) The statistical significance of multivariate Hawkes processes fitted to limit order book data. arXiv:1604.01824

\bibitem[Nelder and Mead(1965)]{Nelder65}
Nelder JA, Mead R (1965) A simplex method for function minimization. Comput J 7:308-313

\bibitem[Panayi et al.(2013)]{Panayi13}
Panayi E, Harman M, Wetherilt A (2013) Agent-based modelling of stock markets using existing order book data. In: Giardini F, Amblard F (eds) Multi-Agent-Based Simulation XIII. Springer-Verlag, Berlin, pp 101-114

\bibitem[Preis et al.(2006)]{Preis06}
Preis T, Golke S, Paul W, Schneider (2006) Multi-agent-based Order Book Model of financial markets. Europhys Lett 75(3):510-516

\bibitem[Recchioni et al.(2015)]{Recchioni15}
Recchioni MC, Tedeschi G, Gallegati M (2015) A calibration procedure for analyzing stock price dynamics in an agent-based framework. J Econ Dyn Control 60:1-25

\bibitem[Rigdon(1995)]{Rigdon95}
Rigdon EE (1995). A necessary and sufficient identification rule for structural models estimated in practice. Multivariate Behav Res 30(3):359-383

\bibitem[Thomson Reuters(2016)]{Reuters16}
Thomson Reuters (2016) Thomson Reuters Tick History. https://tickhistory.thomsonreuters.com. Accessed 23 May 2016

\bibitem[Toke(2011)]{Toke11}
Toke IM (2011) Market Making in an Order Book Model and Its Impact on the Spread. In: Abergel F, Chakrabarti BK, Chakraborti A, Mitra M (eds) Econophysics of Order-driven Markets. Springer, Milan, pp 49-64

\bibitem[Tukey(1977)]{Tukey77}
Tukey JW (1977) Exploratory Data Analysis. Addison-Wesley, Boston, pp 39-49

\bibitem[Wampler(2015)]{Wampler15}
Wampler D (2015) Fast Data: Big Data Evolved. https://info.lightbend.com/rs/558-NCX-702/images/COLL-white-paper-fast-data-big-data-evolved.pdf. Accessed 4 June 2016

\bibitem[Whitley(1994)]{Whitley94}
Whitley D (1994) A genetic algorithm tutorial. Stat Comput 4(2):65-85

\bibitem[Wilcox and Gebbie(2008)]{Wilcox08}
Wilcox D, Gebbie T (2008) Serial correlation, periodicity and scaling of eigenmodes in an emerging market. Int J Theoretical Appl Finance 11(7):739-760

\bibitem[Wilcox and Gebbie(2014)]{Wilcox14}
Wilcox D, Gebbie T (2014) Hierarchical causality in financial economics. SSRN:2544327

\bibitem[Winker et al(2007)]{Winker07}
Winker P, Gilli M, Jeleskovic V (2007) An Objective Function for Simulation Based Inference on Exchange Rate Data. SSRN:964131

\end{thebibliography}
\end{document}